\documentclass[11pt] {article}
\usepackage{epsfig,bm}
\newcommand {\rd}  {{\rm d}}
\newcommand {\re}  {{\rm e}}
\newcommand {\Prob}  {{\rm Prob}}
\newcommand {\FaR}  {{\rm FaR}}

\newcommand {\be}  {\begin{equation}}
\newcommand {\ee}  {\end{equation}}
\newcommand {\bea} {\begin{eqnarray}}
\newcommand {\eea} {\end{eqnarray}}
\newcommand {\nn} {\nonumber}

\begin{document} 
\title{Derivatives and Credit Contagion in Interconnected Networks}
\author{Sebastian Heise\footnote{Present address: Department of Economics, 
Yale University, PO Box 208268, New Haven CT 06520-8268, USA} ~and Reimer K\"uhn\\
Mathematics Department, King's College London, Strand, London WC2R 2LS,UK}
\date{31 Jan 2012}
\maketitle
\vspace{-10mm}
\begin{abstract}
The importance of adequately modeling credit risk has once again been highlighted in the recent financial crisis. Defaults tend to cluster around times of economic stress due to poor macro-economic conditions, {\em but also\/} by directly triggering each other through contagion.  Although credit default swaps have radically altered the dynamics of contagion for more than a decade, models quantifying their impact on systemic risk are still missing. Here, we examine contagion through credit default swaps in a stylized economic network of corporates and financial institutions. We analyse such a system using a stochastic setting, which allows us to exploit limit theorems to exactly solve the contagion dynamics for the entire system. Our analysis shows that, by creating additional contagion channels, CDS can actually lead to greater instability of the entire network in times of economic stress. This is particularly pronounced when CDS are used by banks  to expand their loan books (arguing that CDS would offload the additional risks from their balance sheets). Thus, even with complete hedging through CDS, a significant loan book expansion can lead to considerably enhanced probabilities for the occurrence of very large losses and very high default rates in the system. Our approach adds a new dimension to research on credit contagion, and could feed into a rational underpinning of an improved regulatory framework for credit derivatives.
\end{abstract}

\section{Introduction}

The spectacular growth of the global financial system in the past two decades, followed by its near collapse in 2008, has prompted renewed efforts to assess the risks that may be hidden in the world-wide network of interconnected financial exposures \cite{Haldane09,Turner+10, ECB09, Haldane11}. Yet, although the rapid global expansion of financial markets  has in no small part been driven by a significant increase in derivatives trading, both in volume and in complexity, theoretical investigations into the contribution of derivatives to {\em systemic\/} risk, which would match those  dealing with credit risk at a portfolio or economy-wide level \cite{Lando98, Duffie99, Davis99, Jarrow01, Li01, Frey03, Rogge03, Giesecke04, NeuKu04, HaKu06,Egloff07, HaKu09} are curiously missing.

Credit events tend to cluster in times of economic stress, \cite{Keenan00} forcing banks to recognize disproportionately many  defaults in recessions. This is due to two reasons. First, the profitability of firms depends on common macro-economic variables, such as economic growth, leading to an increase in default rates when macro-economic conditions are poor. Second, firms are directly linked with each other through business relations. The default of a large customer or supplier, for instance, will adversely affect the credit position of a firm, which may then default and in turn influence {\em its\/} customers and suppliers. Such a direct dependency of defaults is referred to as {\em credit contagion}. Both mechanisms underlying clustering of defaults have received considerable attention in the credit risk literature for more than a decade; see \cite{Lando98, Duffie99, Davis99}, and \cite{Davis99, Jarrow01, Li01, Frey03, Rogge03, Giesecke04, NeuKu04, HaKu06,Egloff07, HaKu09}. Evidence suggests that the dependence on common factors can by itself {\em not\/} explain observed levels of correlation \cite{Das07}, and that credit contagion, possibly in conjunction with the effect of further unobserved macro-economic covariates --- so-called frailty --- is important to explain the data \cite{Duff+09, Az+11}.

An entirely new dimension has been added to contagion dynamics in the last two decades through the emergence of credit default swap (CDS) markets, which have created pervasive new forms of financial dependencies. However, while contagion has been addressed in models dealing with the {\em pricing\/} of these instruments  (see e.g. \cite{Hull00, Hull01, Haworth07, Haworth08, Errais10, Frey08, Brigo09, Frey10, Cous+11})). their role as {\em transmitters\/} of contagion and thus as sources of {\em systemic risk\/} has not received much attention. Two recent exceptions deserve mention, viz. \cite{Jorion07}, who takes a first step by looking at CDS spreads as {\em indicators\/} of contagion, and more importantly and closer to our main points of concern \cite{mark+10}, an empirical study which reconstructs the network of financial exposures of major financial players in the US, {\em including\/} exposures due to CDS. A series of stress tests performed by these authors on the reconstructed network of financial exposures clearly demonstrates the destabilizing potential of CDS. We will return to the findings of this paper in relation to ours in greater detail in our final concluding section. 

There have been considerable efforts in recent years to assess levels of risk in the financial system from a networks perspective (e.g., \cite{Elsinger+06, Schwei+09, Cont+09, GaiKap10, Amini+10} and references therein). These have largely concentrated on analysing networks of mutual financial exposures described by pair-interactions. However, as we shall see below, the analysis of contagion dynamics in networks with CDS contracts requires the introduction of networks with hyper-edges corresponding to `three-particle interactions', which have been missing in conventional network approaches. We note that a completely different perspective on stability of financial systems is taken in \cite{Caccioli+09} which describes erosion of {\em market\/} stability resulting from a {\em proliferation of financial instruments\/}.

A CDS is a contract in which a {\em protection buyer} pays a periodic fee (the {\em CDS premium}) to a {\em protection seller}, to protect itself against a potential loss on an exposure to an individual loan or a bond as a result of an unforeseen event (see Fig.1, left panel). Such a {\em credit event} is generally triggered if the {\em reference entity\/} on which the loan or bond was written has become unable to pay interest or principal on its debt. If a credit event occurs, the protection seller has to make a payment to the buyer of the CDS to compensate him for the loss. A CDS is therefore similar to an insurance contract: it provides protection against an adverse event in exchange for a periodic fee. Remarkably, a significant fraction of CDS markets is of a different, {\em speculative\/} nature; in speculative CDS contracts protection buyers need {\em not\/} actually hold a risky exposure on the reference entity (Fig.1, right panel).

\begin{figure}[h!]
\setlength{\unitlength}{1mm}
\begin{picture}(65,40)(0,0)
\put(5,0){\epsfig{file=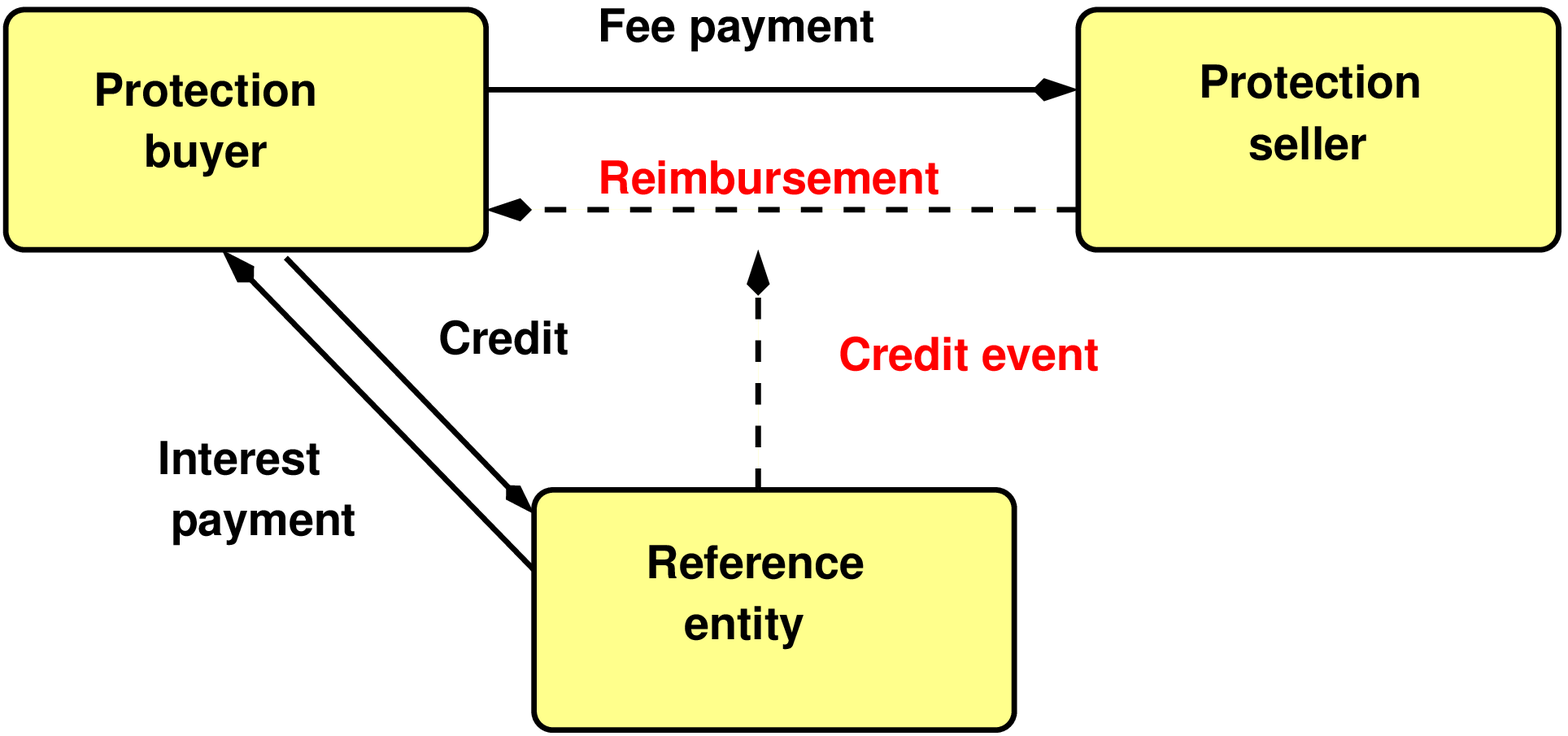,width=7.5cm}}
\put(90,0){\epsfig{file=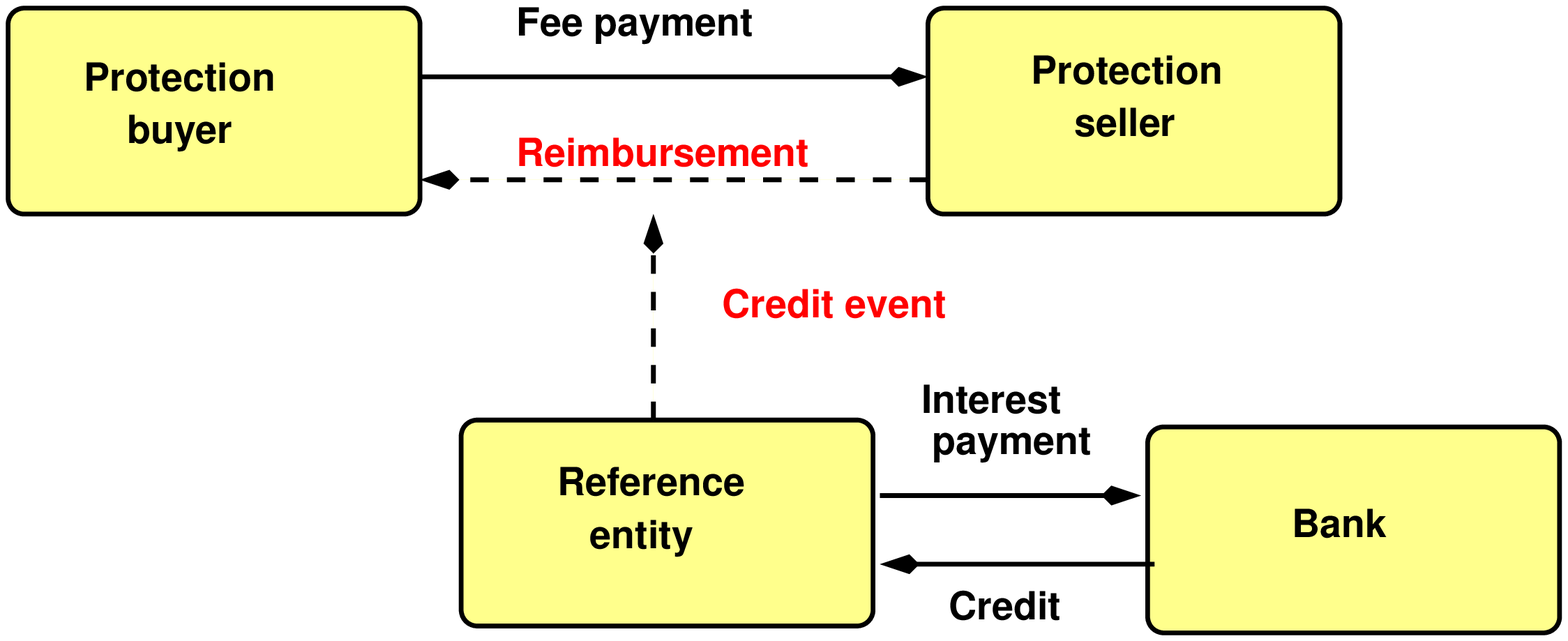,width=8.6cm}}
\end{picture}
\caption{Left: Mechanics of a CDS contract used for hedging. Right: Mechanics 
of a speculative CDS.}
\label{fig:CDSmech}
\end{figure}

According to surveys of the International Swaps and Derivatives Association, the global volume of the CDS market has steadily grown from a value of $\$ 918$\,bn in 2001 to a peak value of $\$ 62$\,trn in 2007, with a subsequent sharp decline in the wake of the financial crisis to $\$ 38$\,trn in 2008, and further down to $\$ 30$\,trn in 2009 \cite{ISDA10}. 

While in theory CDS should reduce credit risk, the recent financial crisis has shown that CDS can be a source of risk in themselves. A case in point is the insurer AIG, which faced bankruptcy after the failure of Lehman Brothers due to the large payouts it was required to make on its CDS contracts referencing Lehman. A default of AIG would have forced {\em its\/} counter-parties to recognize large and unexpected losses on assets they had assumed would not be a source of risk, and would have triggered an avalanche of further defaults across the financial system. These are precisely the dynamics of contagion --- prevented from unfolding only by a costly government bail out \cite{ECB09}.

In view of the significant volume of CDS markets and their well documented prominent role in the unfolding of the recent financial crisis \cite{ECB09} we are thus led to analyze contagion dynamics and systemic risk in networks of financial dependencies which include exposures created by CDS contracts as additional contagion channels. 

We will be looking at an economy composed of interacting heterogeneous networks of `firms' and `financial institutions'. We assume that financial institutions provide credit lines to firms, and that they engage in CDS contracts among each other, either for the purpose of hedging their exposures to firms, or for speculative purposes.

To describe the evolution of an interacting network of wealth positions, we set up a discrete-time stochastic process, in which the probability of default of a given entity (a node in this network) at a particular time step depends on the current, and as we shall see, also the past states of its economic partners, as well as on common macro-economic variables. We restrict our attention at a time horizon $T$ of one year, divided into twelve discrete time steps $\Delta t \equiv 1$, each representing a month. 

The model considered here is a direct generalization of a model first proposed in \cite{NeuKu04} to analyze the influence of economic interactions on credit risk. The two main additional features are {\bf (i)} the presence of three-particle interactions created by CDS contracts, and {\bf (ii)} the non-Markovian nature of contagion effects in systems with CDS contracts. The stochastic characterization of the economic network, and the heuristic arguments used to solve the macroscopic system dynamics follow lines of reasoning introduced earlier to analyse dynamics of neural networks \cite{Derrida87}, and recently applied in analytic studies of contagion dynamics in \cite{HaKu06, HaKu09}. As shown in \cite{HaKu06}, one can invoke the generating functional techniques of \cite{Dominicis78} to demonstrate that the heuristic reasoning is exact in the large system limit. We shall not repeat those rather involved arguments in the present case, however.

The remainder of this document is structured as follows. In Sect. \ref{sec:wealthposdyn} we describe a stochastic default dynamics of the nodes in the network in terms of wealth positions. By analysing wealth positions and the effect of economic interactions on losses for each node in Sect. \ref{sec:Loss}  the default dynamics is shown to capture contagion effects. For financial institutions these include in particular contagion effects from `three-particle' interactions created by CDS contracts. We proceed to analyze contagion in a synthetic probabilistic setting specified in Sect. \ref{sec:network}, which sets up networks of financial exposures as sparse Erd\"os Renyi random graphs \cite{Bollobas01}, and their analogues including hyper-edges that randomly link three nodes through CDS contracts. The macroscopic system dynamics is analysed in Sect. \ref{sec:macrodyn} using the law of large numbers and the central limit theorem to formulate it in terms of a coupled dynamics of the time dependent fractions of defaulted nodes in each sector. Our system is set up in such a way that the statistics of losses generated by defaults is Gaussian, and Sect. \ref{sec:LossT} shows how to express losses in terms of fractions of defaulted nodes in each sector. We provide a detailed system-specification as well as parameter settings for the various scenarios that we have studied in Sect. \ref{sec:Sys}. Sect. \ref{sec:Res} presents our main results in the form of distributions of end-of-year losses {\em per bank\/} and default rates within the banking sector for various scenarios with and without CDS. We conclude in Sect. \ref{sec:Sum} with a summary and discussion. Some technical calculations needed for the analysis of the macroscopic system dynamics are relegated to an appendix.

\section{Wealth Positions and Default Dynamics}
\label{sec:wealthposdyn}

Generalizing \cite{NeuKu04}, we introduce a heterogeneous economic network comprising firms ($F$) and financial institutions. We consider non-financial corporates in this network as counter-parties to loans and bonds, or as reference entities in CDS. The default of a firm will have an economic impact on other firms interacting with it. For the purposes of our analysis we will divide the financial sector into banks ($B$) and insurers ($I$). We reserve for banks the role of directly lending to firms, but assume that both banks and insurers engage in CDS contracts, with insurers typically acting as protection sellers. Thus, we use the tag ``insurer" in a wide sense, to denote financial institutions other than banks acting as counter-parties in CDS contracts.

We describe contagion dynamics in terms of a co-evolution of a network of wealth positions. To this end, we set up a discrete-time stochastic process, in which the probability of default of a node in this network depends on the current and past states of its economic partners, as well as on common macro-economic variables. We restrict our attention to a time horizon $T$ of one year, divided into twelve discrete time steps $\Delta t$.

We denote the wealth position of node $i$ at time $t$ by $W_{i,t}$. A node will default at time $t+\Delta t$ if its wealth position at time $t$ falls below zero. We use indicator variables $n_{i,t}$ to record default, and set $n_{i,t}=1$ ($n_{i,t} = 0$) if node $i$ is in the defaulted (non-defaulted) state at time $t$. Assuming that recovery from default does not occur within the time-frame of a year, we get a default dynamics of the form
\be
n_{i,t+\Delta t} = n_{i,t} + (1-n_{i,t})\Theta\left(-W_{i,t}\right)\ ,
\label{nDyn}
\ee
where $\Theta$ is the Heaviside step function ($\Theta(x)=0$ for $x\le 0$, and $\Theta(x)=1$ for $x > 0$). We follow \cite{NeuKu04, HaKu06, HaKu09}, and take the wealth to be of the 
form
\be
W_{i,t} =\vartheta_i -L_{i,t} + \eta_{i,t}\ , 
\label{Wgen}
\ee
with $\vartheta_i$ denoting the wealth of $i$ at the start of the risk horizon and $L_{i,t}$ the {\em losses\/} accrued up to time $t$ due to defaults of nodes interacting with $i$. The $\eta_{i,t}$ in (\ref{Wgen}) denote fluctuating contributions to the wealth and are taken to be zero-mean Gaussians. In the present study we adopt a minimal noise-model  proposed in the Basel II document, \cite{Basel05} assuming a combination of economy-wide and idiosyncratic components of the form
\be
\eta_{i,t}= \sigma_i\left(\sqrt{\rho_i}\,\xi_{0,t} + \sqrt{1-\rho_i}\, \xi_{i,t}\right)\ .
\label{eta}
\ee
In (\ref{eta}), the parameter $\rho_i$ describes the correlation of the noise $\eta_{i,t}$ and its macro-economic component $\xi_{0,t}$, and $\sigma_i$ is the standard deviation of the total noise. The Basel II document suggests to link the correlation factors $\rho_i$ with unconditional (annual) default probabilities ${\rm PD}_i$ according to
\be
\rho_i \simeq 0.12 \left(1+ \re^{-50 {\rm PD}_i} \right)\ . 
\label{BaselII}
\ee
It is assumed that the $\{\xi_{i,t}\}$ in (\ref{eta}) are independent in $i$ and $t$, and that the economy-wide noise varies {\em slowly\/} in comparison with the idiosyncratic noise. To simplify matters, we will assume $\xi_{0,t}$ to be {\em constant\/} throughout the period of the risk horizon $T$ of one year, $\xi_{0,t}\equiv \xi_{0}$, and obtain {\em loss-distributions\/} by looking at how annual losses vary with $\xi_{0}$.

\subsection{Contagion and Losses}
\label{sec:Loss}
A complete characterization of the contagion dynamics requires specifying the dependence of the losses $L_{i,t}$ on the set of indicator variables $\{n_{i,t}\}$.

For {\em firms\/} we assume losses to be generated through defaults of other nodes directly interacting with $i$, given by
\be
L_{i,t} = L^{(d)}_{i,t} =\sum_{j} J^{(d)}_{ij}\, n_{j,t}\ , \qquad i\in F \ .
\label{Lfirm}
\ee
The coupling $J^{(d)}_{ij}$ in (\ref{Lfirm}) quantifies the material impact on the wealth of firm $i$ that would be caused by a default of node $j$. Depending on whether $j$ has a cooperative or a competitive relation with $i$, a default of $j$  may result in a loss ($J^{(d)}_{ij} > 0$) or a gain ($J^{(d)}_{ij} < 0$). Clearly $J^{(d)}_{ij}=0$, if there is no direct interaction between $i$ and $j$.

We assume that {\em financial institutions\/} engage in several types $\alpha$ of interaction among each other and with firms, with losses $L^{(\alpha)}_{i,t}$ depending on the type $\alpha$ of interaction. The losses could derive from direct economic interactions ($d$) as for firms, from exposures due to unhedged loans ($u$), due to hedged loans where the financial institution takes the role of either the protection buyer ($hb$), or protection seller ($hs$), and finally from speculative buying ($sb$) or selling ($ss$) of CDS. Thus losses of a financial institution due to defaults within the network of its business partners take the form
\be
L_{i,t} = \sum_\alpha L^{(\alpha)}_{i,t}\  ,\qquad i\in B \cup I \ ,
\label{Lfin}
\ee
with loss types $L^{(\alpha)}_{i,t}$ given as follows.

\begin{itemize}
\item Losses due to direct economic interactions with defaulting entities are defined as for firms:
\be
L^{(d)}_{i,t} = \sum_{j} J_{ij}^{(d)}\, n_{j,t}
\label{Ld}
\ee
Direct economic interactions could well cross sector boundaries, as firms may provide services to banks or insurers, banks and insurers could share accounting and payment systems, and so on.

\item Unhedged loans generate losses through defaults, but also an income stream from interest payments received from debtors while they are still alive. For a loan of size $J^{(u)}_{ij}$ given to $j$ they amount to $J_{ij}^{(u)} \epsilon_{ij,\tau}$ 
in month $\tau$, $1\le \tau \le t$:
\be
L^{(u)}_{i,t}= \sum_{j\in F,B} J_{ij}^{(u)}\, \sum_{\tau=1}^t\, \Big[(n_{j,\tau} 
-n_{j,\tau-1}) - \epsilon_{ij,\tau}\Big]
\label{Lus}
\ee
The aggregation of monthly losses through contagion constitutes a telescoping sum, simplifying to $\sum_{\tau=1}^t\, (n_{j,\tau} - n_{j,\tau-1}) = n_{j,t}$. Interest paid up to time $t$ depends on the entire history of each debtor. We will specify $\epsilon_{ij,\tau}$ below.

\item Hedged loans generate losses through defaults of reference entity $j$ occurring at a time where the protection seller $k$ is also in a defaulted state. Protection fee payments also contribute to losses; they are exchanged only while all three counter-parties involved are still alive, and amount to $J^k_{ij} f_{ij,\tau}^k$ at time $\tau$. Hedged loans also generate an income stream from interest payments, as long as the reference entity is alive:
\be
L^{(hb)}_{i,t}=\sum_{j\in F,B}\sum_{k\in B,I} J_{ij}^k\, \sum_{\tau=1}^t \left[ 
(n_{j,\tau}-n_{j,\tau-1}) n_{k,\tau} + f_{ij,\tau}^k- \epsilon_{ij\tau}\right]
\label{Lhbs}
\ee
In (\ref{Lhbs}), $J^k_{ij}$ denotes the exposure of $i$ to the reference entity $j$, which is hedged through a CDS contract with $k \in B \cup I$; both fee payments and interest income are proportional to the exposure. We will specify $f_{ij,\tau}^k$, the amount of fees paid per (hedged) unit loan below. {\bf Note:} Losses through default of the reference entity are incurred only if the protection seller is dead {\em at time of default\/}. If the protection seller defaults only later, compensation payments will have been made, and losses will not have been incurred.

\item A protection seller $i$ for hedged loans, while still alive, incurs losses through default of reference entity $j$, and derives an income stream from regular fee payments as long as all counter-parties involved are still alive:
\be
L^{(hs)}_{i,t}= \sum_{j\in F,B}\sum_{k\in B} J_{ij}^k\, \sum_{\tau=1}^t\, \left[(n_{j,\tau} 
-n_{j,\tau-1})(1-n_{i,\tau}) -  f_{ij,\tau}^k\right] 
\label{Lhs}
\ee

\item Speculative protection buyers derive income from credit events, provided the protection seller $k$ is still alive when the credit event occurs. They incur losses through regular fee payments as long as all counter-parties involved are still alive. Denoting the size of the speculative exposure by $K_{ij}^k$, we have
\be
L^{(sb)}_{i,t}= - \sum_{j\in F,B}\sum_{k\in B,I} K_{ij}^k\,\sum_{\tau=1}^t\, 
\left[(n_{j,\tau} -n_{j,\tau-1})(1- n_{k,\tau}) -  f_{ij,\tau}^k \right]
\label{Lsb}
\ee

\item Speculative protection sellers, while still alive, incur losses from credit events. They derive income from regular fee payments, as long as all counter-parties involved are still alive:
\be
L^{(ss)}_{i,t} = \sum_{j\in F,B}\sum_{k\in B} K_{ij}^k\, \sum_{\tau=1}^t\,
\left[(n_{j,t}-n_{j,\tau-1})(1-n_{i\tau}) -  f_{ij,\tau}^k\right]
\label{Lss}
\ee
\end{itemize}

Note that there are no self-interactions in (\ref{Lfirm})--(\ref{Lss}), i.e., for all $i$ we have that $J_{ii}=0$, and $J_{ii}^k=J_{ij}^i=0$ as well as $K_{ii}^k=K_{ij}^i=0$ irrespectively of $j$ or $k$. Also, as CDS contracts have protection buyer and a protection seller taking symmetrically opposite counter-positions, we must have $J_{ij}^k = J_{kj}^i$, and similarly $K_{ij}^k = K_{kj}^i$.  Note also that CDS create {\em three node\/} interactions. Finally, although the structure of losses for speculative and non-speculative protection sellers is basically the same, we keep them separate, because they differ by the nature of the counter-parties involved; they may, and in general also will, exhibit different statistics of exposure sizes and and connectivities.

Due to the history dependence of the fee and interest money-streams and the losses suffered through contagion effects in exposures involving CDS contagion dynamics in our system will in general be {\em non-Markovian.}

\subsection{Interest and Fees}
\label{sec:int-fees}

Interest payments are only received while the entity taking out a loan is still alive. In a similar vein, we assume that a CDS premium is only exchanged as long as all counter-parties involved are still alive, as default of the reference entity triggers compensation payments, and terminates the contract, while default of the protection seller effectively removes protection and thus the obligation to pay fees (for non-existing protection), and finally default of the protection buyer makes the protection buyer unable to pay for protection.

To keep matters simple, we ignore capital repayments and use interest rates that are kept constant throughout the risk horizon of a year. If $\epsilon_{ij}$ denotes the monthly fraction of the debt paid in interests by $j$ to $i$, while $j$ is still alive, interest payments in month $\tau$ on a unit loan are given by
\be
\epsilon_{ij,\tau} = \epsilon_{ij} (1+\epsilon_{ij})^{\tau - 1}(1-n_{j,\tau})\ .
\label{eq:int}
\ee
Hence, if $j$ survives up to and only up to time $t^* \le t$, then the total interest payed is $e_{ij,t} = \sum_{\tau=1}^t \epsilon_{ij,\tau} = (1 + \epsilon_{ij})^{t^*} -1$ for all $t\ge t^*$.

For the results presented in the present study we simplify matters further by assuming that interest rates only depend on the sectors of counter-parties, $\epsilon_{ij} = \epsilon_{ss'}$, with $s=s(i)$ and $s'=s'(j)$. Our simplifying assumptions could easily be relaxed, but are not thought to be crucial for our main purpose of assessing losses from contagion dynamics, and in particular the tail region of high losses.

As to fees, we assume that fee sizes are heterogeneous, and fluctuate over time to capture the fact that protection buyers may engage in offsetting contracts, if the credit quality of a counter-party changes. We use $f_{ij,\tau}^k$ to denote the fee per unit exposure paid in month $\tau$ in a CDS contract between protection  buyer $i$ and protection seller $k$, with $j$ as reference entity, and take it to be of the form
\be
f_{ij,\tau}^k = (f_0 + f \, y_{ij,\tau}^k)(1-n_{i,\tau})(1-n_{j,\tau})
(1-n_{k,\tau}) \ .
\label{eq:fee}
\ee
I.e. fees are paid only as long as all three counter-parties involved are still alive Here the term $f_0$ represents the average monthly fee paid in the economy, while the $y_{ij,\tau}^k$, taken to be zero-mean, unit variance Gaussians, are introduced to capture the fluctuations of CDS premia across entities and over time, with $f$ describing the scale of these fluctuations. We assume the $y_{ij,\tau}^k$ to be identically distributed and independent in $i,j,k$ and $\tau$.

\subsection{Probabilistic Characterization of the Economic Network}
\label{sec:network}

\begin{figure}[t]
$$\epsfig{file=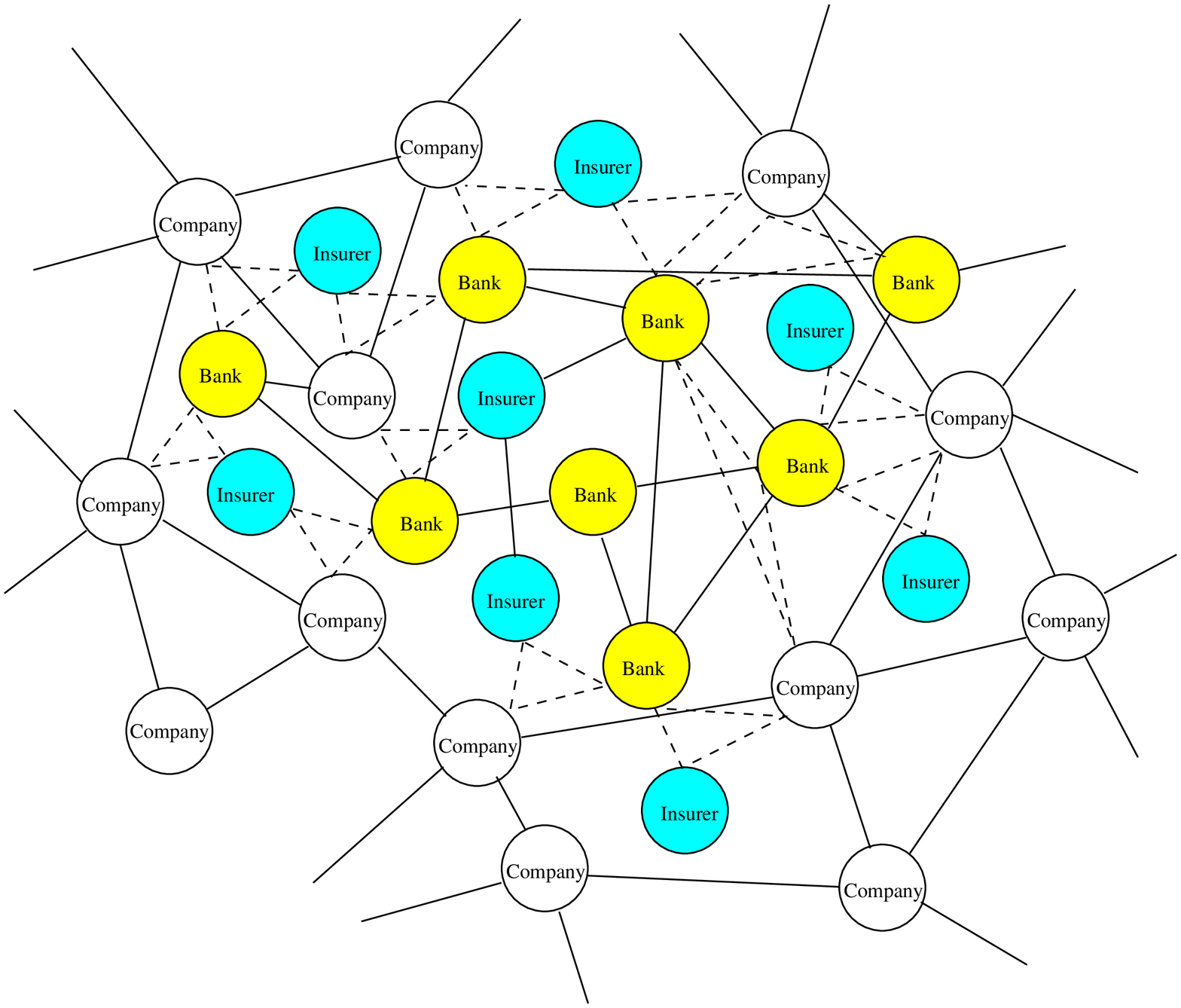, width=7cm}$$
\caption{Schematic representation of a modular financial network of firms, banks and insurers. Full lines represent direct interactions or exposures due to unhedged loans, dashed triangles represent three-particle interactions created through CDS.} 
\label{fig:FinNet}
\end{figure}

We study a synthetic version of the problem using a stochastic setting in which we assume weighted random graph structures for the interconnected networks of mutual exposures of firms $F$, banks $B$, and insurers $I$, as shown schematically in Figure \ref{fig:FinNet}. We use  $N_F$, $N_B$, and $N_I$ to denote the numbers of firms, banks, and insurers, respectively, and we assume each of these numbers to be large. 

We take exposures due to direct interaction or from unhedged loans to be  of the form
\be
J^{(\alpha)}_{ij} = c^{(\alpha)}_{ij} \left(\frac{\bar 
J^{(\alpha)}_{rs}}{C^{(\alpha)}_{rs}} + \frac{J^{(\alpha)}_{rs}}
{\sqrt{C^{(\alpha)}_{rs}}}~ x^{(\alpha)}_{ij} 
\right)\ ,
\label{Jdir}
\ee
in which $\alpha\in\{d,u\}$, and $r=r(i)$ and $s=s(j)$ denote the sectors to which the counter-parties belong ($r,s \in \{F,B,I\}$), and $C^{(\alpha)}_{rs}$ the average number of connections of type $\alpha$ of a node in sector $r$ to nodes in sector $s$. In (\ref{Jdir}), we isolated connectivities $c^{(\alpha)}_{ij} \in \{0,1\}$ which indicate the presence ($c^{(\alpha)}_{ij} =1$) or absence ($c^{(\alpha)}_{ij} =0$) of an exposure of $i$ to $j$ from its size. The connectivities are taken to be symmetric, $c^{(\alpha)}_{ij} =c^{(\alpha)}_{ji}$, and chosen independently in pairs according to the probability distribution 
\be
 P(c^{(\alpha)}_{ij}) = \frac{C^{(\alpha)}_{rs}}{N_s} 
\delta_{c^{(\alpha)}_{ij},1} + \left(1-\frac{C^{(\alpha)}_{rs}}{N_s}\right)
\delta_{c^{(\alpha)}_{ij},0}\  .
\label{Pcij}
\ee
In the limit of large sector sizes at fixed $C^{(\alpha)}_{rs}$, this specification generates networks of mutual exposures which form a modular Erd\"os-Renyi random graph \cite{Bollobas01}, exhibiting Poissonian degree distributions with on average $C^{(\alpha)}_{r,s}$ links of type $\alpha$ from nodes in sector $r$ to nodes in sector $s$, where $r,s\in \{F,B,I\}$.

The $x^{(\alpha)}_{ij}$  characterizing individual exposure sizes in (\ref{Jdir}) are pair-wise independent random variables, with low-order moments given by
\be
\overline {x_{ij}^{(\alpha)}} =0  \  ,\quad \overline {(x^{(\alpha)}_{ij})^2}=1\  ,
\quad \mbox{and} \quad \overline {x^{(\alpha)}_{ij} x^{(\alpha)}_{ji}}=\kappa^{(\alpha)}_{rs}\ .
\label{xij}
\ee
Thus $\bar J^{(\alpha)}_{rs}$ and $J^{(\alpha)}_{rs}$ parameterize the mean and standard deviation of type-$\alpha$ exposures (if present) of nodes in sector $r$ to nodes in sector $s$, and $\kappa^{(\alpha)}_{rs}$ their forward-backward correlation. We are interested in the limit where $N_{s} \gg 1$ for all $s$, and $C^{(\alpha)}_{rs} \gg 1$ for all $r,s$; connectivities are also assumed to be sparse in the sense that $C^{(\alpha)}_{rs} \ll N_s$. As already noted in \cite{HaKu06, HaKu09}, it is a remarkable consequence of the absorbing nature of the defaulted state that the macroscopic system dynamics will be independent of the degree of forward-backward correlations paremeterized by the $\kappa^{(\alpha)}_{rs}$.

Analogous specifications are used to describe exposures resulting from CDS contracts which involve {\em three\/} nodes of the entire network, the protection buyer $i$, the reference entity $j$, and the protection seller $k$. With reference to Eqs. (\ref{Lhbs}) and (\ref{Lsb}), the specification from the protection buyer's perspective reads
\be
J_{ij}^k = c_{ij}^k \left(\frac{\bar J_{b,r}^s}{C_{b,r}^s} 
+ \frac{J_{b,r}^s}{\sqrt{C_{b,r}^s}}~ x_{ij}^k \right)
\label{Jcds}
\ee
with $b=b(i)$, $r=r(j)$, $s=s(k)$ denoting the sectors of protection buyer, reference entity and protection seller, respectively. Here $C_{b,r}^s$ denotes the average number of CDS contracts with reference entities in $r$ and protection sellers in $s$ a buyer (in $b$) is engaged in. As in (\ref{Jdir}), we have isolated information about the existence of contracts from information about their size. We choose the $c_{ij}^k$ and the $x_{ij}^k$ to be independent with
\be
 P(c_{ij}^k) = \frac{C_{b,r}^s}{N_{r}N_{s}} \delta_{c_{ij}^k,1} 
+ \left(1-\frac{C_{b,r}^s}{N_{r}N_{s}}\right) 
\delta_{c_{ij}^k,0}\ ,
\label{Pcijk}
\ee
and the first two moments of the $x_{ij}^k$ to be given by
\be
\overline{x_{ij}^k}=0  \  ,\quad \mbox{and} \quad \overline{(x_{ij}^k)^2}=1\  ,
\label{xijk}
\ee
so that $\bar J_{b,r}^s$ and $J_{b,r}^s$ parameterise mean and standard deviation of CDS exposures. As CDS contracts have protection buyer and a protection seller taking symmetrically opposite counter-positions, we must have $J_{ij}^k = J_{kj}^i$, or equivalently $c_{ij}^k= c_{kj}^i$ and $x_{ij}^k = x_{kj}^i$. We assume that the average numbers of CDS contracts financial institution are engaged in are large, $C_{b,r}^s \gg 1$, with connectivities sparse in the sense that $C_{b,r}^s \ll N_r N_s$. 
Fully analogous conventions are used to characterise the statistics of exposures $K^k_{ij}$ in speculative CDS contracts.

The sparseness and large average connectivity assumptions used in the present paper are mainly made for the sake of analytic tractability: they are needed to allow a relatively straightforward heuristic solution of the macroscopic system dynamics, as described in the following section.  Our solution uses methods originally devised to study the dynamics of sparsely connected neural network \cite{Derrida87}. As shown in  \cite{HaKu06}, some of the decorrelation assumptions used along the way can be fully justified by recourse to generating functional methods. We note in passing that the scaling of means and variances of interaction parameters with mean connectivities in Eqs. (\ref{Jdir}) and (\ref{Jcds}) entails, that actual values of these mean connectivities will not appear in final expressions.

\section{Solving the Macroscopic System Dynamics}
\label{sec:macrodyn}

\subsection{Equations of Motion at System-Level}

To analyze the contagion dynamics at system level, we first combine (\ref{nDyn}) and (\ref{Wgen}) to get
\be
n_{it+\Delta t} = n_{i,t} + (1-n_{i,t})\Theta\left(L_{i,t} -\vartheta_i -\eta_{i,t}
\right) .
\label{nDynL}
\ee
Key to the solution is the observation that the losses $L_{i,t}$ appearing in (\ref{nDynL}) contain contributions from {\em large numbers\/} of counter-parties. The contributions to these losses are random due to the heterogeneous specification of the exposures, and due to the fluctuating contributions to the wealth positions. Because of the sparseness of connectivities and the statistical independence of exposure sizes, individual contributions to the losses are sufficiently weakly correlated to entail that the losses $L_{i,t}$ and individual loss types $L_i^{(\alpha)}$ listed in (\ref{Ld}) -- (\ref{Lss}) are Gaussian  by the central limit theorem (CLT). That is,
\be
L_{i,t}^{(\alpha)} = \bar L_{i,t}^{(\alpha)} + \sqrt{V_{i,t}^{(\alpha)}} \, \zeta_{i,t}^\alpha\ , 
\label{LiGauss}
\ee
with $\zeta_{i,t}^\alpha$ i.i.d. Gaussians of zero mean and unit variance. In Sect. \ref{sec:LossT} we will compute the means $\bar L_{i,t}^{(\alpha)}$ and variances $V_{i,t}^{(\alpha)}$ explicitly and show that they depend only on the sector $s$ to which $i$ belongs, thus $\bar L_{i,t}^{(\alpha)}=\bar L_{s,t}^{(\alpha)}$ and $V_{i,t}^{(\alpha)} =V_{s,t}^{(\alpha)}$, and that they are expressible in terms of the fractions
\be
m_{s,\tau} = \frac{1}{N_s} \sum_{i\in s} n_{i,\tau}\ ,\qquad \tau \le t,
\label{msts}
\ee
of companies in the various sectors $s\in \{F,B,I\}$ that have defaulted at various times up to time $t$. 

The non-Markovian nature of the default dynamics is due to the fact that aggregate fee and interest payments as well as contagion-losses in CDS contracts depend on the entire history of the counter-parties involved. For a general non-Markovian stochastic dynamics, one would typically expect that an analysis of the macroscopic system dynamics would necessitate the 
evaluation of correlation functions, i.e. of quantities depending on two time arguments, which would in turn usually require deriving of equations of motion for them. It is a remarkable fact that this is not needed in the present case, a simplification which is entirely due to the absorbing nature of the defaulted state in our system, entailing that the required 
correlation functions have simple expressions in terms of one-time quantities.

In what follows we will denote averages w.r.t. the fluctuating wealth contributions by angled brackets, and averages over the heterogeneous composition of the economic network by an over-bar. Performing these averages in (\ref{nDynL}) we get
\be
\overline {\langle n_{it+\Delta t}\rangle} = \overline {\langle n_{it}\rangle} 
+ \overline {\langle (1-n_{i,t})\Theta\left(L_{i,t} -\vartheta_i -\eta_{i,t}\ ,
\right)\rangle} 
\ee
which describes the evolution of the {\em probability\/} $\overline {\langle n_{it} \rangle}$ of node $i$ to have defaulted. Evaluating the average in the second term in terms of a joint average over the independent the $n_{i,t}$, $L_{i,t}$ and $\xi_{i,t}$ distributions \cite{HaKu06, HaKu09}, keeping the economy-wide noise $\xi_{0,t}$ in the noise specification (\ref{eta}) fixed, and exploiting the fact that both $L_{i,t}$ and the idiosyncratic noise $\xi_{i,t}$ in (\ref{eta}) are Gaussian. we get
\be
\overline {\langle n_{it+\Delta t}\rangle} = \overline {\langle n_{it}\rangle} 
+ \Big(1-\overline {\langle n_{i,t}\rangle}\,\Big) \Phi\left(\frac{\bar L_{s,t} - \sigma_i 
\sqrt{\rho_i}\,\xi_{0,t} - \vartheta_i} {\sqrt{ V_{s,t} + \sigma_i^2(1-\rho_i)}}\right)
\label{nav}
\ee
in which $\Phi$ is the integrated normal density
\be
\Phi(x) = \int_{-\infty}^x \frac{\rd z}{\sqrt{2\pi}}\, \re^{-z^2/2}\ .
\ee
Note that the probability $\overline {\langle n_{it}\rangle}$ of node $i$ to have defaulted depends on $i$ {\em only\/} through the initial wealth $\vartheta_i$, the standard-deviation $\sigma_i$ of the noise, and the correlation parameter 
$\rho_i$. We will use the notation $\langle n_{t}(\sigma_i, \vartheta_i, \rho_i)\rangle \equiv \overline {\langle n_{it}\rangle}$ to highlight this fact. In terms of this convention then, Eq. (\ref{nav}) for nodes in $s$ with 
$(\sigma_i,\vartheta_i, \rho_i) = (\sigma,\vartheta, \rho)$ reads
\be
\langle n_{t+\Delta t}(\sigma,\vartheta,\rho)\rangle = \langle n_{t}(\sigma,
\vartheta,\rho)\rangle +\Big(1- \big\langle n_{t}(\sigma,\vartheta,\rho)
\big\rangle \Big) \,
\Phi\left(\frac{\bar L_{s,t} - \sigma \sqrt{\rho}\,\xi_{0,t} - \vartheta} 
{\sqrt{ V_{s,t} + \sigma^2(1-\rho)}}\right)\ .
\label{nav2}
\ee
Eq. (\ref{nav2}) has a simple intuitive interpretation: as the defaulted state is assumed to be absorbing, the probability of nodes in $s$ with $(\sigma,\vartheta,\rho)$ to be defaulted at time $t+\Delta t$ is given by the probability that they were defaulted at time $t$, plus the  conditional probability that they would default at time $t+\Delta t$, given the system 
state and given that they were not defaulted at time $t$, multiplied by by probability that they were indeed not defaulted at time $t$.

The macroscopic system dynamics is obtained by looking at $m_{s,t+\Delta t}$, using (\ref{nDynL}) and (\ref{msts}),
\be
m_{s,t+\Delta t} = \frac{1}{N_s} \sum_{i\in s} \Big[ n_{i,t}  +  (1-n_{i,t})
\Theta\left(L_{i,t} -\vartheta_i -\eta_{i,t} \right) \Big]\ ,
\label{mdyn1}
\ee
and evaluating the sum (\ref{mdyn1}) as a sum of averages by appeal to the law of large numbers. With averages from (\ref{nav}) or (\ref{nav2}) we get 
\be
m_{s,t+\Delta t} = m_{s,t} + \frac{1}{N_s} \sum_{i\in s} \Big(1- \big\langle 
n_{t}(\sigma_i,\vartheta_i, \rho_i)\big\rangle\,\Big) \Phi\left(\frac{\bar L_{s,t} - 
\sigma_i \sqrt{\rho_i}\,\xi_{0,t} - \vartheta_i} {\sqrt{ V_{s,t} + 
\sigma_i^2(1-\rho_i)}}\right)\ .
\label{mdyn2}
\ee
We can appeal once more to the law of large numbers to express the sum in (\ref{mdyn2}) as an average over the joint $(\sigma, \vartheta, \rho)$-distribution, finally giving 
\be
m_{s,t+\Delta t} = m_{s,t} + \left\langle \Big(1- \big\langle 
n_{t}(\sigma,\vartheta,\rho)\big\rangle\,\Big) \Phi\left(\frac{\bar L_{s,t} - 
\sigma \sqrt{\rho}\,\xi_{0,t} - \vartheta} {\sqrt{ V_{s,t} + 
\sigma^2(1-\rho)}}\right)\right\rangle_{\sigma,\vartheta,\rho}\ ,
\label{mdyn3}
\ee
where $\langle \dots\rangle_{\sigma,\vartheta,\rho}$ denotes the average over the $(\sigma,\vartheta, \rho)$-distribution. Once we have expressed means and variance of losses in terms of fractions $m_{s,t}$ of defaulted nodes in each sector, Eq. (\ref{mdyn3}) describes the macroscopic dynamics of the system fully in terms of the coupled dynamics of the $m_{s,t}$.

\subsection{Statistics of Loss-Types}
\label{sec:LossT}

Below we explicitly compute mean and variance for losses of firms or financial institutions due to direct interactions, and for losses of banks due to unhedged loans. The computation for the other loss types follows along the same lines, and we will only list the results. The calculation of the various contributions to loss-variances in particular requires the evaluation of correlations of indicator variables for different time arguments; these calculations  are performed in Appendix \ref{sec:AppCorr}; they exploit the absorbing nature of the defaulted state and ultimately allow loss-variances to be expressed in terms of one-time quantities. To simplify notation in the following argument we omit the superscript $(d)$ signifying the direct loss channel from parameters characterizing mean and standard deviation of exposure-sizes and average connectivities.

Consider direct losses of $i$, which could be a firm or a financial institution, due to direct interactions. Denote by $s=s(i)$ the sector of $i$. Inserting the specification of the $J^{(d)}_{ij}$ from (\ref{Jdir}) into the expression for direct losses, we have
\bea
L_{i,t}^{(d)}&=&\sum_{j} J^{(d)}_{ij} n_{j,t} = \sum_{s'}\sum_{j\in s'} J^{(d)}_{ij} n_{j,t}
= \sum_{s'}\Bigg[\frac{\bar J_{ss'}}{C_{ss'}}\sum_{j\in s'} c_{ij} n_{j,t} +
\frac{J_{ss'}}{\sqrt{C_{ss'}}} \sum_{j\in s'} c_{ij}~ x_{ij} n_{j,t}\Bigg]\ .
\label{Lddec}
\eea
We obtain for the mean $\bar L_{i,t}^{(d)} \equiv \overline{\langle L_{i,t}^{(d)}\rangle}$
\bea
\bar L_{i,t}^{(d)} &=& \sum_{s'}\Bigg[\frac{\bar J_{ss'}}{C_{ss'}}\sum_{j\in s'} 
\overline{c_{ij} \langle n_{j,t}\rangle} + \frac{J_{ss'}}{\sqrt{C_{ss'}}} 
\sum_{j\in s'} \overline{c_{ij}\,x_{ij} ~ \langle n_{j,t}\rangle}\Bigg]\ .
\eea
Assuming that noise averages and compositional averages are sufficiently weakly correlated that we may factor averages w.r.t. composition, we get
\bea
\bar L_{i,t}^{(d)} &\simeq& \sum_{s'}\Bigg[\frac{\bar J_{ss'}}{C_{ss'}}\sum_{j\in s'}
\overline{c_{ij}}~\overline{\langle n_{j,t}\rangle} + \frac{J_{ss'}}{\sqrt{C_{ss'}}} 
\sum_{j\in s'} \overline{c_{ij}}~\overline{\,x_{ij}}~\overline{\langle n_{j,t}\rangle}\Bigg]
\nn\\
&=& \sum_{s'}\frac{\bar J_{ss'}}{N_{s'}} \sum_{j\in s'}\overline{\langle n_{j,t}\rangle}
= \sum_{s'} \bar J_{ss'} m_{s',t}
\eea
Here we have also exploited independence of the $c_{ij}$ and the $x_{ij}$ and the fact that $\overline{\,x_{ij}}=0$ and $\overline{c_{ij}}=C_{ss'}/N_{s'}$ in view of (\ref{Pcij}), when $i\in s$ and $j\in s'$. The factorisation of averages used in 
this argument can be fully justified in the limit of large system size using the functional techniques in \cite{HaKu06}.

Note that the first contribution to the direct loss (\ref{Lddec}) is equal to the mean by the law of large numbers,
$$
\bar L_{i,t}^{(d)} \simeq \sum_{s'}\frac{\bar J_{ss'}}{C_{ss'}} \sum_{j\in s'} 
c_{ij}~ n_{j,t}\ ,
$$
when evaluated as a sum of averages.

Following the same line of reasoning, we evaluate the variance of $L_{i,t}^{(d)}$ for $i\in s$ as 
\bea
V_{i,t}^{(d)} &\simeq& \sum_{s'}\frac{J_{ss'}^2}{{C_{ss'}}} 
\sum_{j,j'\in s'}\overline{c_{ij}c_{ij'}}~\overline{x_{ij}x_{ij'}} ~
\overline{\langle n_{j,t}n_{j',t}\rangle} = \sum_{s'}\frac{J_{ss'}^2}{{C_{ss'}}} 
\sum_{j\in s'}\overline{c_{ij}}~\overline{\langle n_{j,t}\rangle} \nonumber\\
&=& \sum_{s'}\frac{J_{ss'}^2}{{N_{s'}}}\sum_{j\in s'}~\overline{\langle n_{j,t}\rangle} 
=\sum_{s'} J_{ss'}^2 m_{s',t}\ ,
\eea
where we have exploited independence of the $x_{ij}$, and $n_{j,t}^2=n_{j,t}$

Next we look at losses incurred by $i\in s$ due to unhedged loans given to entities $j\in s'$, as given by Eq. (\ref{Lus}); in the standard situation we would have $s=B$ and $s'=F$. Using analogous factorization properties for averages, as well as 
$\overline{x_{ij}}=0$, and the expression (\ref{eq:int}) for interest payments, we get
\be
\bar L_{it}^{(u)} \simeq \frac{\bar J_{ss'}}{C_{ss'}} \sum_{j\in s'} \overline{c_{ij}} \left[
\overline{\langle n_{j,t}\rangle} - \sum_{\tau=1}^t\overline{\langle \epsilon_{i,j,\tau}\rangle}\right]
=\bar J_{ss'} \left[m_{s',t} - \epsilon_{ss'} \sum_{\tau=1}^t (1+\epsilon_{ss'})^{\tau-1}
(1-m_{s',\tau}) \right]
\ee
The computation of the variance exploits the same factorization and independence properties, giving 
$$
V_{i,t}^{(u)} \simeq \frac{J_{ss'}^2}{N_{s'}} \sum_{j\in s'}\left[\overline{\langle n_{j,t}
\rangle} + \sum_{\tau,\tau'=1}^t \overline{\langle \epsilon_{ij,\tau}
\epsilon_{ij,\tau'}\rangle} -2\sum_{\tau=1}^t \overline{\langle n_{j,t}
\epsilon_{ij,\tau}\rangle}\right]\ .
$$
To proceed, we need
\be
\sum_{\tau,\tau'=1}^t \overline{\langle \epsilon_{ij,\tau}\epsilon_{ij,\tau'}\rangle} 
= \epsilon_{ss'}^2 \sum_{\tau,\tau'=1}^t (1+\epsilon_{ss'})^{\tau-1}
(1+\epsilon_{ss'})^{\tau'-1} \overline{\langle (1-n_{j,\tau})(1-n_{j,\tau'})\rangle}\ ,
\label{e:eps-eps}
\ee
and
\be
\sum_{\tau=1}^t \overline{\langle n_{j,t}\epsilon_{ij,\tau}\rangle} 
= \epsilon_{ss'}\sum_{\tau =1}^t (1+\epsilon_{ss'})^{\tau-1}\,\overline{\langle n_{j,t} 
(1-n_{j,\tau})\rangle}\ ,
\label{e:c-eps}
\ee
which involve two-time correlations of indicator variables. Further evaluation is facilitated by noting that due to the absorbing nature of the defaulted state we have $(1-n_{j,\tau})(1-n_{j,\tau'}) = (1-n_{j,\tau^>})$,  where $\tau^> = {\rm max}
\{\tau,\tau'\}$ and $n_{j,t} (1-n_{j,\tau}) =n_{j,t} -n_{j,\tau}$ for $\tau \le t$, which allows to express the required two-time quantities by reducing them to single-time quantities. Putting things together we then get
\bea
V_{i,t}^{(u)} &=& J_{ss'}^2\left[m_{s',t} +\epsilon_{ss'}\sum_{\tau =1}^t 
(1+\epsilon_{ss'})^{\tau-1}\Big[(1+\epsilon_{ss'})^{\tau-1}+ 
(1+\epsilon_{ss'})^{\tau}-2\Big](1-m_{s',\tau})\right .\nonumber\\
& & ~~~~~~~~~~~~~~ \left . - 2\epsilon_{ss'}\sum_{\tau =1}^t (1+\epsilon_{ss'})^{\tau-1}
(m_{s,t}-m_{s,\tau})\right]
\eea

The computation of loss statistics for the remaining loss types listed in  (\ref{Ld})--(\ref{Lss}) uses analogous factorization and reduction properties. Below we therefore just list general results, referring to Appendix \ref{sec:AppCorr} 
for the calculations used to reduce two-time quantities needed for the evaluation of variances to one-time quantities. Using $s=s(i)$, $s'=s'(j)$, and $s''=s''(k)$ to denote sectors of counter-parties, we get
\begin{itemize}
\item for losses from direct exposures
\be
\bar L_{i,t}^{(d)}= \sum_{s'} \bar J_{s,s'} m_{s',t}
\ee
\be
V_{i,t}^{(d)}= \sum_{s'} J_{s,s'}^2 m_{s',t}
\ee
\item for losses from unhedged loans
\be
\bar L^{(u)}_{i,t} = \sum_{s'} \bar J_{s,s'}\Bigg[m_{s',t} -\epsilon_{s,s'}\sum_{\tau=1}^{t} 
(1+\epsilon_{s,s'})^{\tau-1}(1-m_{s',\tau})\Bigg]
\ee
\bea
V^{(u)}_{i,t} &=& \sum_{s'} J_{s,s'}^2\Bigg[m_{s',t} +\epsilon_{ss'}\sum_{\tau=1}^{t} 
(1+\epsilon_{ss'})^{\tau-1}\Big[(1+\epsilon_{ss'})^{\tau-1}+ (1+\epsilon_{ss'})^{\tau}-2\Big]
(1-m_{s',\tau})\nonumber\\
& & ~~~~~~~~~~ - 2\epsilon_{ss'}\sum_{\tau=1}^{t} (1+\epsilon_{ss'})^{\tau-1}(m_{s',t}-m_{s',\tau})
\Bigg]
\eea
\item for losses from hedged loans (conditioned on $n_{i,t}=0$)
\bea
\bar L^{(hb)}_{i,t}\Big|_{n_{it}=0} &=& \sum_{s',s''} \bar J_{s,s'}^{s''}\sum_{\tau=1}^t 
\Bigg[(m_{s',\tau}- m_{s',\tau-1}) m_{s'',\tau} + f_0 (1-m_{s',\tau})(1-m_{s'',\tau})\nn\\
& &  ~~~~~~~~~~~~~~~~ -\epsilon_{s,s'} (1+\epsilon_{s,s'})^{\tau-1}(1-m_{s',\tau})  \Bigg]
\eea
\bea
V^{(hb)}_{i,t}\Big|_{n_{it}=0} &=& \sum_{s',s''} (J_{s,s'}^{s''})^2 \sum_{\tau=1}^t
\Bigg[(m_{s',\tau}-m_{s',\tau-1}) m_{s'',\tau}\nn\\
& &+  \Big[f_0^2\,(2\tau-1)+f^2 \Big] (1-m_{s',\tau})(1-m_{s'',\tau}) \nn\\
& & +\epsilon_{ss'}(1+\epsilon_{ss'})^{\tau-1}
\Big[(1+\epsilon_{ss'})^{\tau-1}+ (1+\epsilon_{ss'})^{\tau}-2\Big](1-m_{s',\tau})
\nn\\
& & +2 f_0 (m_{s',\tau}-m_{s',\tau-1}) \sum_{\tau'=1}^{\tau}\,(m_{s'',\tau}-m_{s'',\tau'}) 
\nn\\
& & -2 \Big[1+\epsilon_{s,s'})^{\tau-1}-1\Big] (m_{s',\tau}-m_{s',\tau-1}) m_{s'',\tau} 
\nn\\
& & - 2 f_0 \Bigg(\Big[(1+\epsilon_{ss'})^{\tau-1}-1\Big](1-m_{s',\tau})
(1-m_{s'',\tau}) 
\nn\\
& &   ~~~~~~~~~~~~~~~~ + \epsilon_{ss'}(1+\epsilon_{ss'})^{\tau-1}(1-m_{s',\tau})\,
\sum_{\tau'=1}^{\tau} (1-m_{s'',\tau'}) \Bigg)\Bigg]
\eea

\item for losses from protections selling (conditioned on $n_{it}=0$)
\be
\bar L^{(hs)}_{i,t}\Big|_{n_{it}=0} = \sum_{s',s''} \bar J_{s,s'}^{s''}
\Bigg[m_{s',t} - f_0 \sum_{\tau=1}^t (1-m_{s',\tau})(1-m_{s'',\tau})
\Bigg]
\ee

\bea
V^{(hs)}_{i,t}\Big|_{n_{it}=0} &=& \sum_{s',s''} (J_{s,s'}^{s''})^2 \Bigg[m_{s',t}
+ \sum_{\tau=1}^{t} \Big[f_0^2 (2\tau-1)+f^2 \Big] (1-m_{s',\tau})(1-m_{s'',\tau}) \nonumber\\
& & ~~~~~~~~~~~ -2 f_0 \sum_{\tau=1}^{t}\sum_{\tau'=1}^{\tau-1}(m_{s',\tau}-m_{s',\tau-1})
(1-m_{s'',\tau'})\Bigg]
\eea
\item for losses from speculative protection buying (conditioned on $n_{it}=0$)
\be
\bar L^{(sb)}_{i,t}\Big|_{n_{it}=0} = - \sum_{s',s''} \bar K_{s,s'}^{s''} \sum_{\tau=1}^{t} 
\Bigg[(m_{s',\tau}-m_{s',\tau-1}) (1-m_{s'',\tau})
-f_0 (1-m_{s',\tau})(1-m_{s'',\tau})\Bigg]
\label{e:Lsb}
\ee
\bea
V^{(sb)}_{i,t}\Big|_{n_{it}=0} &=& \sum_{s',s''} (K_{s,s'}^{s''})^2  
\sum_{\tau=1}^{t} \Bigg[(m_{s',\tau}-m_{s',\tau-1})(1-m_{s'',\tau}) \nn\\
& &  ~~~~~~~~~~~ + \Big[f_0^2 (2\tau-1)+f^2 \Big] (1-m_{s',\tau})(1-m_{s'',\tau}) \nn\\
& & ~~~~~~~~~~~ -2 f_0 (m_{s',\tau}-m_{s',\tau-1})(1-m_{s'',\tau})\,(\tau-1) \Bigg]
\eea
\item for losses from speculative protection selling  (conditioned on $n_{it}=0$)
\be
\bar L^{(ss)}_{i,t}\Big|_{n_{it}=0} = \sum_{s',s''} \bar K_{s,s'}^{s''}\Bigg[m_{s',t}
- f_0 \sum_{\tau=1}^{t} (1-m_{s',\tau})(1-m_{s'',\tau})\Bigg]
\label{e:Lss}
\ee
\bea
V^{(ss)}_{i,t}\Big|_{n_{it}=0} &=& \sum_{s',s''} (K_{s,s'}^{s''})^2 \Bigg[m_{s',t} 
+ \sum_{\tau=1}^{t} \Big[f_0^2 (2\tau-1)+f^2 \Big] (1-m_{s',\tau})(1-m_{s'',\tau})\nn\\
& & ~~~~~~~~~~~~~~~ -2 f_0 \sum_{\tau=1}^{t}(m_{s',t}-m_{s',\tau})(1-m_{s'',\tau})\Bigg]
\eea
\end{itemize}
Note that means and variances of losses incurred by node $i$ depend only on the sector $s=s(i)$ to which node $i$ belongs, and on the fractions of defaults in the various sectors.

\section{System and Scenario Specifications}
\label{sec:Sys}
 
\subsection{System Specifications}

For the purposes of the present study we adopt system specifications to reflect generic properties of heterogeneous networks of economic players. Some of our assumptions, notably those concerning the nature of mutual connectivities of various economic players described in Sect. \ref{sec:network} above, are even largely dictated by the demands of analytic tractability. While the present synthetic approach can thus by no means claim to be microscopically realistic, we believe that our main message, as well as qualitative trends concerning the contribution of CDS to systemic risk, which we extract from a {\em comparison of scenarios with and without CDS}, holds tight and will survive (re-)introducing further microscopic detail, through calibration or otherwise.

We assume that initial wealths $\vartheta_i$ are randomly distributed in each sector, with sector dependent statistics. Specifically, we take the $\vartheta_i$ to be normally distributed with means depending on the sector, $\bar \vartheta_F=2.75$, $\bar \vartheta_B=3.25$ and $\bar \vartheta_I=3.75$, but standard deviation $\sigma_\vartheta =0.35$ independently of the sector. If we take standard deviations $\sigma_i$ of the noise (\ref{eta}) to be uniform throughout the system, $\sigma_i\equiv 1$, this amounts to assuming heterogeneous unconditional monthly default probabilities $p_i =\Phi(-\vartheta_i)$ \cite{NeuKu04}. With the sector means of the $\vartheta_i$ as specified above, we obtain sector-specific typical monthly default probabilities, which are in the range $10^{-6} - 10^{-2}$ for the firm sector, with values roughly down by a factor 50 for the banking sector, and by another factor 50 for the insurance sector.

\subsection{Scenario Specifications}
\label{sec:Scen}

We specify {\em scenarios\/} by choosing values of the vectors $\bm J_{ss'}^{(\alpha)} = (\bar J_{ss'}^{(\alpha)}, J_{ss'}^{(\alpha)})$ which parametererize mean and variance of exposure size distributions for interactions of type $\alpha$ between sectors $s$ and $s'$ via Eq. (\ref{Jdir}), as well as the $\bm J_{br}^{s} = (\bar J_{br}^{s}, J_{br}^{s})$ and $\bm K_{br}^{s} = (\bar K_{br}^{s}, K_{br}^{s})$ which encode this information for the statistics of non-speculative and speculative CDS exposures involving nodes in sectors $b$, $r$, and $s$ via Eq. (\ref{Jcds}).

We define a {\em baseline-scenario\/} by considering a network consisting of firms and banks only, choosing  $\bm J^{(d)}_{FF} = (1,1)$ and $\bm J^{(d)}_{BB} =(0,0.5)$ for direct interactions within the firm and banking sectors, and $\bm J^{(u)}_{BF} =(1,0.5)$ for unhedged lending by banks to firms. We assume that there are no CDS contracts in the  baseline-scenario. Other scenarios will be compared with the baseline. In all scenarios we keep $\bm J^{(d)}_{FF}$ and $\bm J^{(d)}_{BB}$ unchanged, and we choose monthly interest to be at 0.5\% uniformly across sectors. In scenarios with CDS, we choose parameters describing average and variance of CDS spreads in Eq. (S.14) as $f_0=0.0008$ (amounting to an annual average spread of 1\%, or 100 bp), and $f^2=0.0002$, respectively. 

The following list contains parameter settings for the different scenario's considered in the present project, inasmuch as they change between scenarios.

\begin{description}
\item[$\bm S_0$] baseline scenario: $\bm J^{(u)}_{BF} =(1,0.5)$ 
\item[$\bm S_1$] doubling unhedged lending to firms: $\bm J^{(u)}_{BF} =(2,1)$
\item[$\bm S_2$] doubling unhedged lending by adding inter-bank lending: $\bm J^{(u)}_{BF} =\bm J^{(u)}_{BB} =(1,0.5)$.
\item[$\bm S_3$] hedging one third of base-line exposures with banks: $\bm J^{(u)}_{BF} =(0.67,0.33)$, $\bm J^{B}_{BF} =(0.33,0.17)$
\item[$\bm S_4$] hedging two thirds of base-line exposures with banks: $\bm J^{(u)}_{BF} =(0.33,0.17)$, $\bm J^{B}_{BF} =(0.67,0.33)$
\item[$\bm S_5$] hedging one third of base-line exposures with insurers: $\bm J^{(u)}_{BF} =(0.67,0.33)$, $\bm J^{I}_{BF} =(0.33,0.17)$
\item[$\bm S_6$] hedging two thirds of base-line exposures with insurers: $\bm J^{(u)}_{BF} =(0.33,0.17)$, $\bm J^{I}_{BF} =(0.67,0.33)$
\item[$\bm S_7$] totally hedged exposure doubled in size compared to base-line, all additional exposure hedged with insurers: $\bm J^{(u)}_{BF} =(0,0)$, $\bm J^{B}_{BF} =(0.5,0.25)$, $\bm J^{I}_{BF} =(1.5,0.75)$
\item[$\bm S_8$] totally hedged exposure tripled in size compared to base-line, all additional exposure hedged with insurers: $\bm J^{(u)}_{BF} =(0,0)$, $\bm J^{B}_{BF} =(0.5,0.25)$, $\bm J^{I}_{BF} =(2.5,1.25)$
\item[$\bm S_{9}$] adding speculative CDS of a volume matching that of the base-line exposure (banking sector only): $\bm J^{(u)}_{BF} =(1,0.5)$, $\bm K^{B}_{BF} =(1,0.5)$
\item[$\bm S_{10}$] adding speculative CDS of a volume twice that of the base-line exposure (banking sector only): $\bm J^{(u)}_{BF} =(1,0.5)$, $\bm K^{B}_{BF} =(2,1)$
\item[$\bm S_{11}$] adding speculative CDS in equal measure with banks and insurers: $\bm J^{(u)}_{BF} =(1,0.5)$, $\bm K^B_{BF}=\bm K^I_{BF}=(0.25,0.125)$:
\item[$\bm S_{12}$] adding speculative CDS in equal measure with banks and insurers: $\bm J^{(u)}_{BF} =(1,0.5)$, $\bm K^B_{BF}=\bm K^I_{BF}=(0.5,0.25)$
\item[$\tilde{\bm S}_{0}$] modified baseline scenario with non-zero average for direct inter-bank exposures: $\bm J^{(d)}_{BB} = (0.25,0.5)$, keeping all other parameters fixed as in the baseline scenario.
\item[$\tilde{\bm S}_{9}$] modified version of $\bm S_{9}$ with a non-zero average for direct inter-bank exposures $\bm J^{(d)}_{BB} = (0.25,0.5)$, keeping all other parameters fixed as in $\bm S_{9}$.
\item[$\tilde{\bm S}_{10}$] modified version of $\bm S_{10}$ with a non-zero average for direct inter-bank exposures $\bm J^{(d)}_{BB} = (0.25,0.5)$, keeping all other parameters fixed as in $\bm S_{10}$.
\end{description}

\section{Results}
\label{sec:Res}

We proceed to illustrate contagion and the statistics of losses in our system, both with and without CDS. We concentrate 
on end-of-year losses {\em per node\/} in the banking sector, given by $L = \bar L_{B,T} = \frac{1}{N_B} \sum_{i\in B} L_{iT}$, and on the defaulted fraction of banks $m = m_{B,T} = \frac{1}{N_B} \sum_{i\in B} n_{iT}$. It is worth emphasising the difference between the distribution of individual losses $L_{iT}$ --- shown to be Gaussian --- and the distribution of losses per bank as displayed in Figs. 3-7 below. As the banking sector is taken to be large, the losses per bank are given by the {\em average\/} loss per bank in the banking sector by the law of large numbers. This average itself depends on the random variable $\xi_0$ which parameterizes macro-economic conditions. Although the $\xi_0$-distribution was itself assumed to be Gaussian, it generates distributions of losses per bank displayed in Figs 3-7 below, which are manifestly {\em non-Gaussian\/}. This is due to the non-linear feedback through contagion dynamics that exists in the 
system.

\begin{figure}[t!]
\setlength{\unitlength}{1mm}
\begin{picture}(165,60)(0,0)
\put(5,00){\epsfig{file=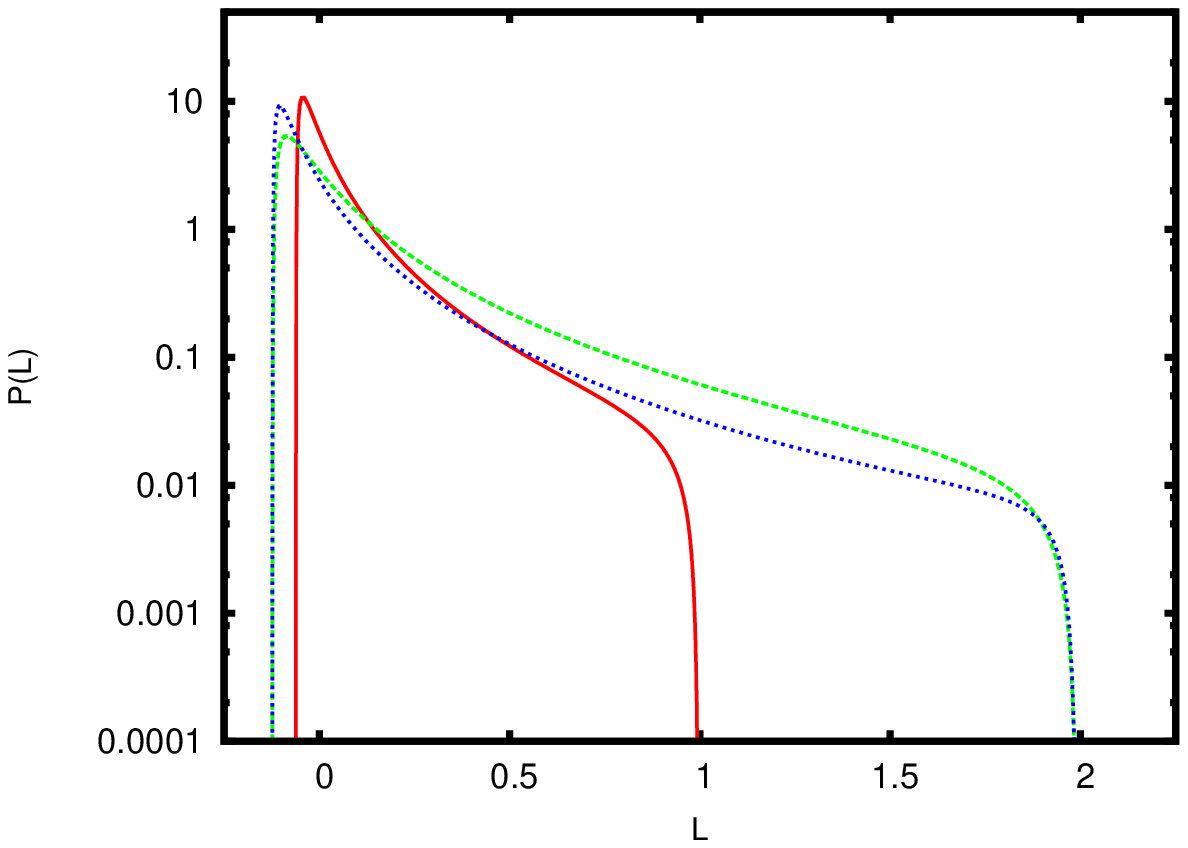,width=8.0cm}}
\put(90,00){\epsfig{file=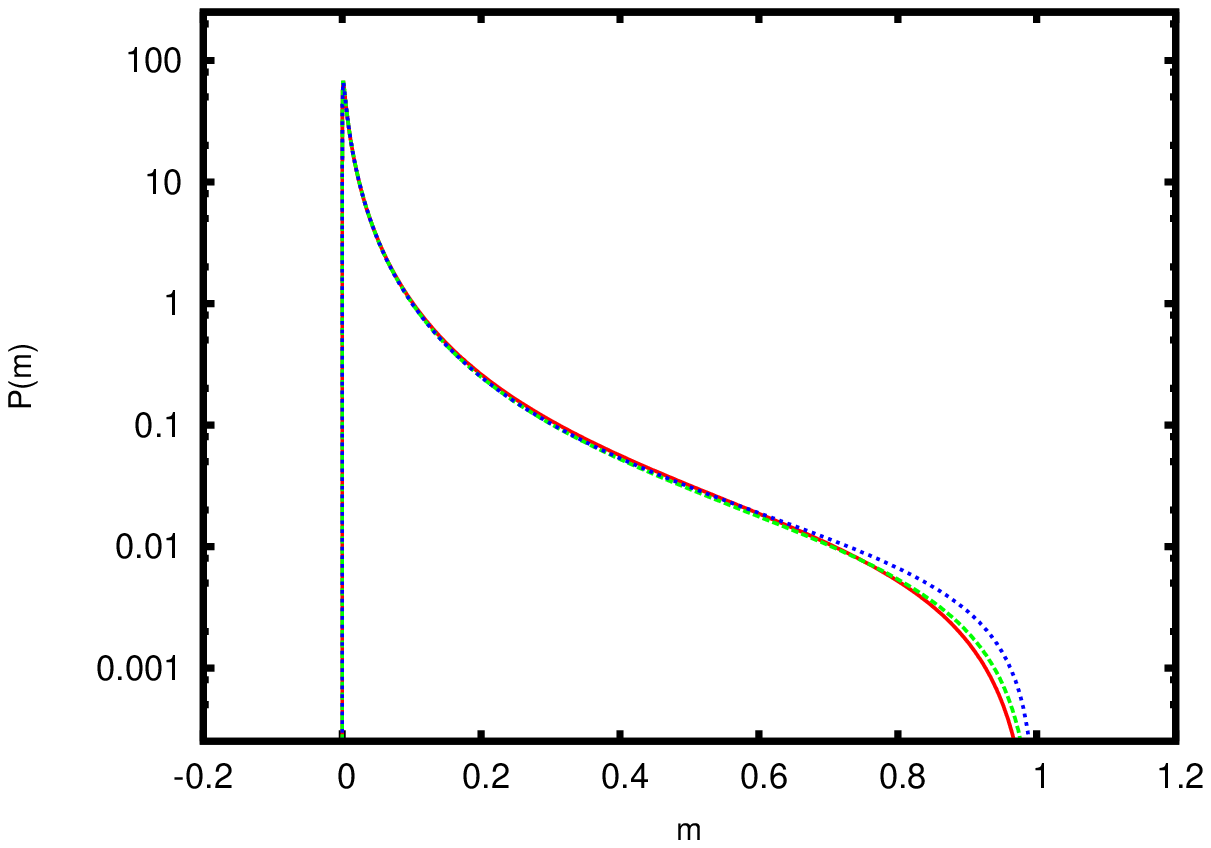,width=8.0cm}}
\put(129,27.5){\epsfig{file=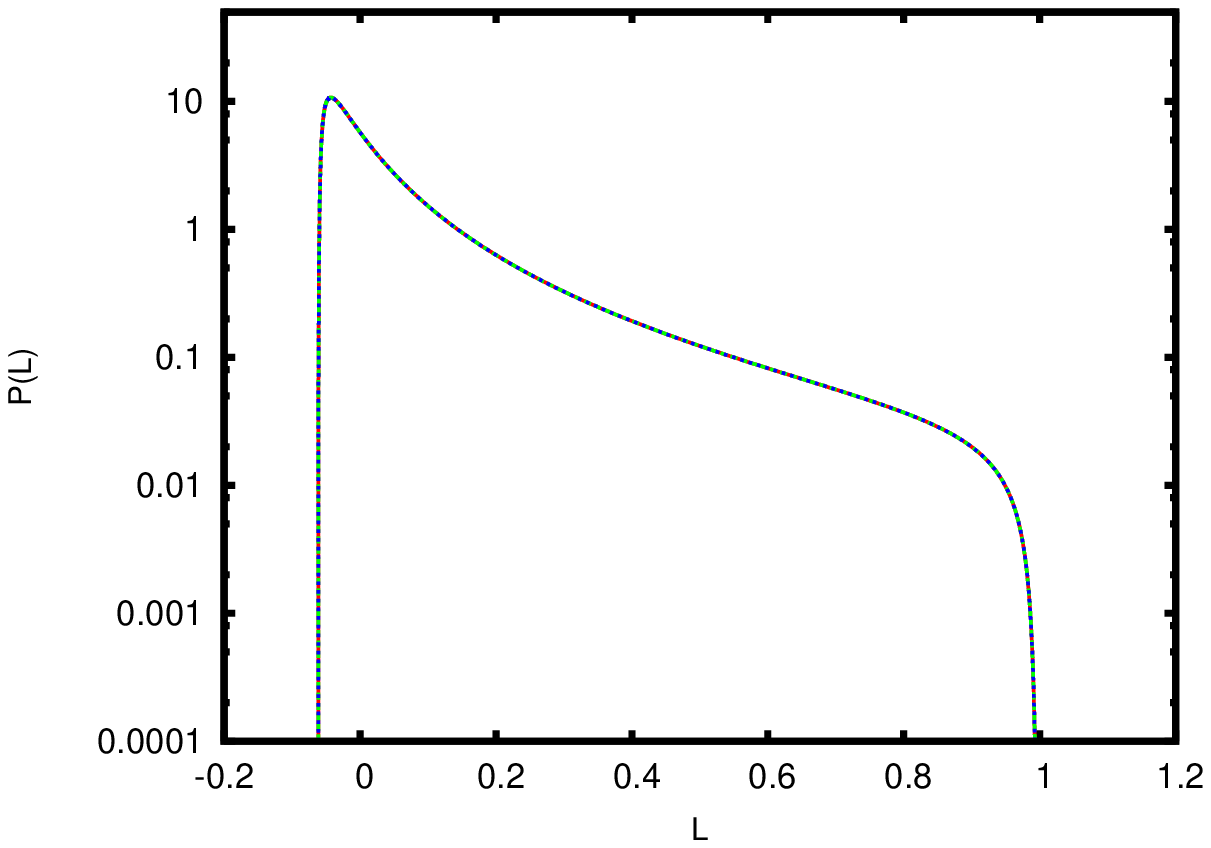,width=3.5cm}}
\end{picture}
\caption{Left: Loss distributions for the baseline-scenario ($\bm S_0$, red full line), compared to scenarios with increased exposures: doubling average exposures by choosing $\bm J^{(u)}_{BF}=(2,1)$ ($\bm S_1$, green long-dashed), or $\bm J^{(u)}_{BF}=\bm J^{(u)}_{BB} = (1,0.5)$ ($\bm S_2$, blue short-dashed). Right: Distribution of the fraction of defaulted banks in the baseline scenario ($\bm S_0$, red), compared to situations where either one third ($\bm S_3$, green long-dashed) or two thirds ($\bm S_4$, blue short-dashed) of the average baseline-exposure to firms are hedged by CDS {\em within the banking sector\/}. The corresponding loss curves are shown in the inset, and they lie exactly on top of each other, as hedging via CDS is modelled as a zero-sum game in our description.}
\label{fig:Res1ab}
\end{figure}

In what fallows we will discuss our results by way of comparison with those pertaining to the {\em baseline-scenario\/} $\bm S_0$ without CDS, consisting of a network of firms and banks only, with direct interactions within the firm and banking sectors and levels of unhedged lending by banks to firms as specified above.

As shown in Fig. \ref{fig:Res1ab}, doubling the size of bank's loan books will approximately double their typical earning (the location of the maximum of the loss distribution at negative losses), but will also double the size of maximal losses. The second panel of Fig. \ref{fig:Res1ab} compares the baseline-scenario with a situation where either one third or two thirds of the baseline exposures of banks to firms are hedged by CDS {\em within the banking sector\/}. The main panel shows that the additional contagion channels created by CDS lead to an increase of probabilities for large fractions of defaults, {\em although the overall exposure to firms remained unchanged in these scenarios}. The corresponding loss curves are shown in the inset, and they lie exactly on top of each other, as hedging via CDS is modelled as a zero-sum game in our description.

\begin{figure}[t!]
\setlength{\unitlength}{1mm}
\begin{picture}(165,60)(0,0)
\put(5,00){\epsfig{file=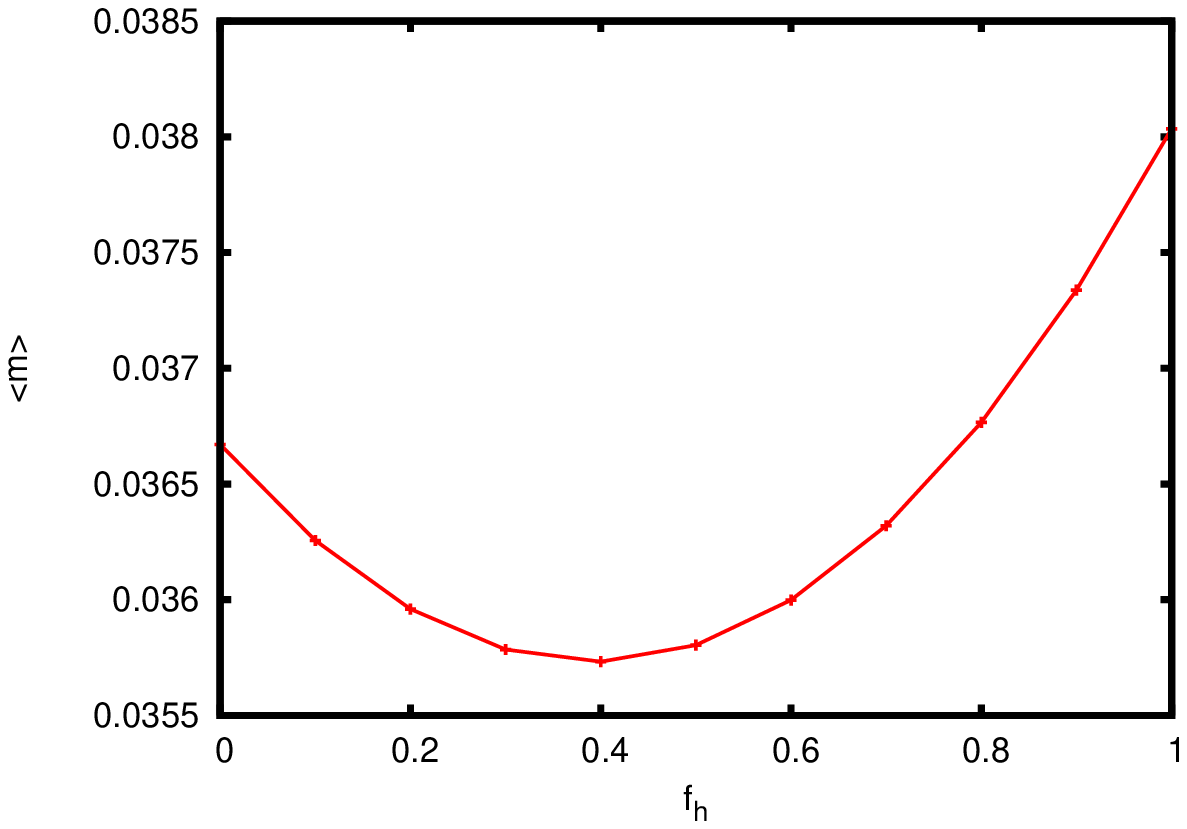,width=8.0cm}}
\put(25,24){\epsfig{file=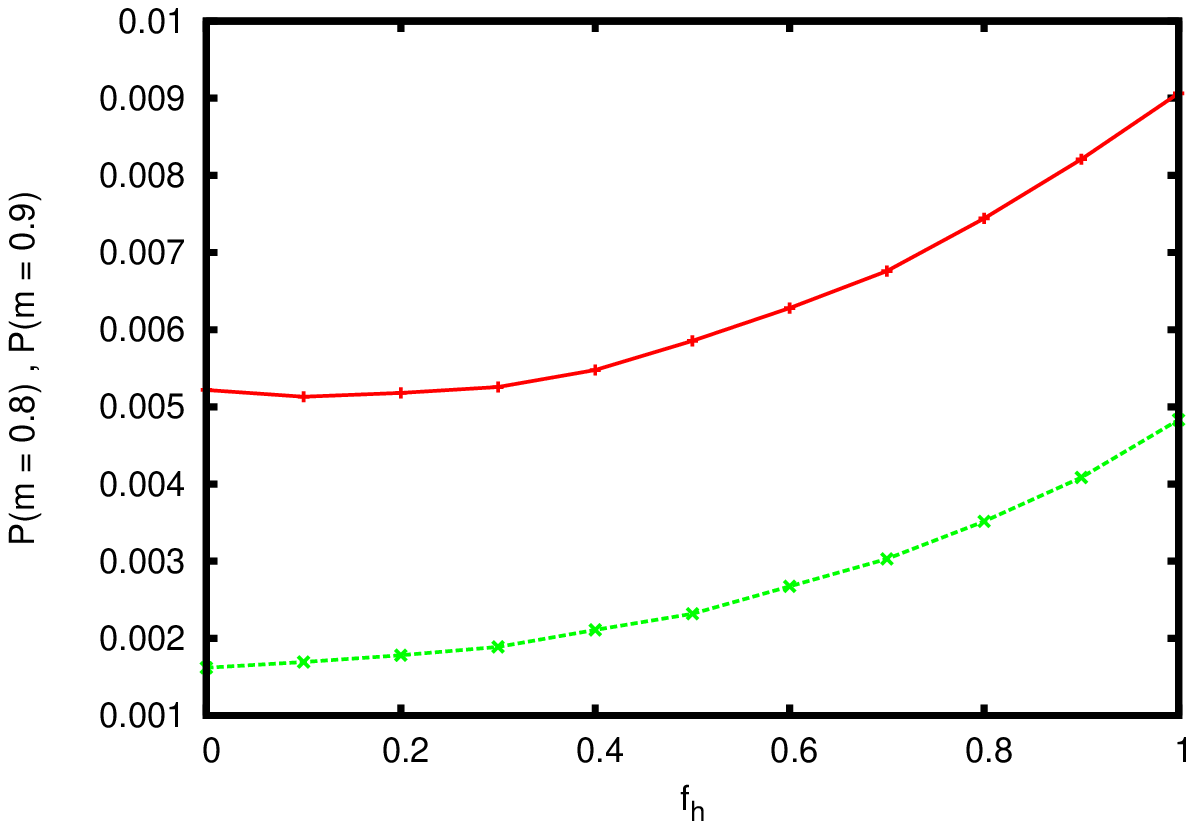,width=4cm}}
\put(90,0){\epsfig{file=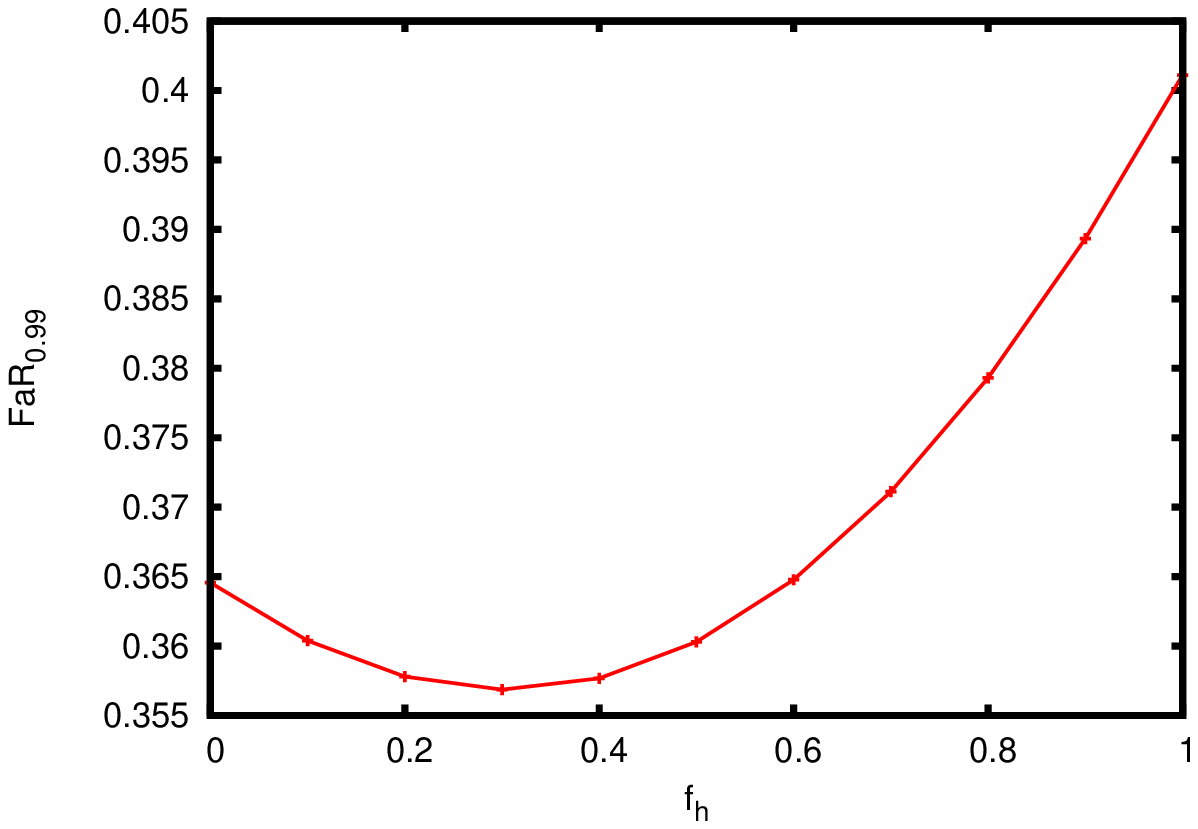,width=8.0cm}}
\end{picture}
\caption{Left: Average fraction $\langle m\rangle$ of defaulted banks as a function of the fraction $f_h$ of the baseline-scenario exposure to firms that is hedged within the banking sector, the remainder remaining unhedged. The inset shows the value of the probability density function $P(m)$ evaluated at $m=0.8$ (upper curve) and $m=0.9$ (lower curve) as a function of $f_h$. Right: Fraction at Risk at the 99\% confidence level as a function of the fraction of baseline-scenario exposure to firms hedged within the banking sector}
\label{fig:Res1cd}
\end{figure}

Figure \ref{fig:Res1cd} investigates the effect of hedging inside the banking sector more systematically. Starting from 
the baseline scenario, we look at the behaviour of a set of risk-measures as a function of the fraction $f_h$ of baseline exposures of banks to firms which are hedged {\em inside the banking sector}, the remaining fraction remaining unhedged. The left panel shows the average fraction $\langle m\rangle = \int \rd m~ m P(m)$ of defaulted banks. This average fraction initially decreases as a function of $f_h$, but starts to rise again as $f_h$ is increased beyond $f_h \simeq 0.4$, with the average fraction in the fully hedged situation being more than 3.5\% higher than that in the fully unhedged baseline scenario! Looking at the probability density function $P(m)$ of the fraction of defaulted banks at a given level $m$, it shows a monotone increase as a function of $f_h$ at large $m$, rising to almost twice the baseline level at full hedging $f_h=1$ for $m=0.8$, with only a minute initial drop, and to three times the baseline level at full hedging for $m=0.9$.

Alternatively, and in close analogy to the Value at Risk (VaR) concept, we are led to introduce the notion of Fraction at Risk ($\FaR_q$), by which we denote the fraction of failed nodes in excess of the average fraction, which is not exceeded at a given confidence level $q$. Specifically, given a confidence level $q$, one defines the $q$-quantile $m_q$ of $m$ via $\Prob\{m\le m_q\}=q$, and in terms of these
\be
\FaR_q = m_q - \langle m\rangle\ .
\ee
The right panel of Fig \ref{fig:Res1cd} exhibits the behaviour of $\FaR_q$ at the 99\% confidence level as a function  of the fraction $f_h$ of baseline exposures to firms which is hedged inside the banking sector. Its behaviour is qualitatively similar to that of the average fraction $\langle m\rangle$, exhibiting a minimum at a slightly smaller value $f_h \simeq 0.3$, and rising to a level which at full hedging is 10\% {\em above\/} baseline level. Note once more that the loss distribution is {\em independent\/} of the level of hedging within the banking sector.

\begin{figure}[t!]
\setlength{\unitlength}{1mm}
\begin{picture}(165,62)(0,0)
\put(5,0){\epsfig{file=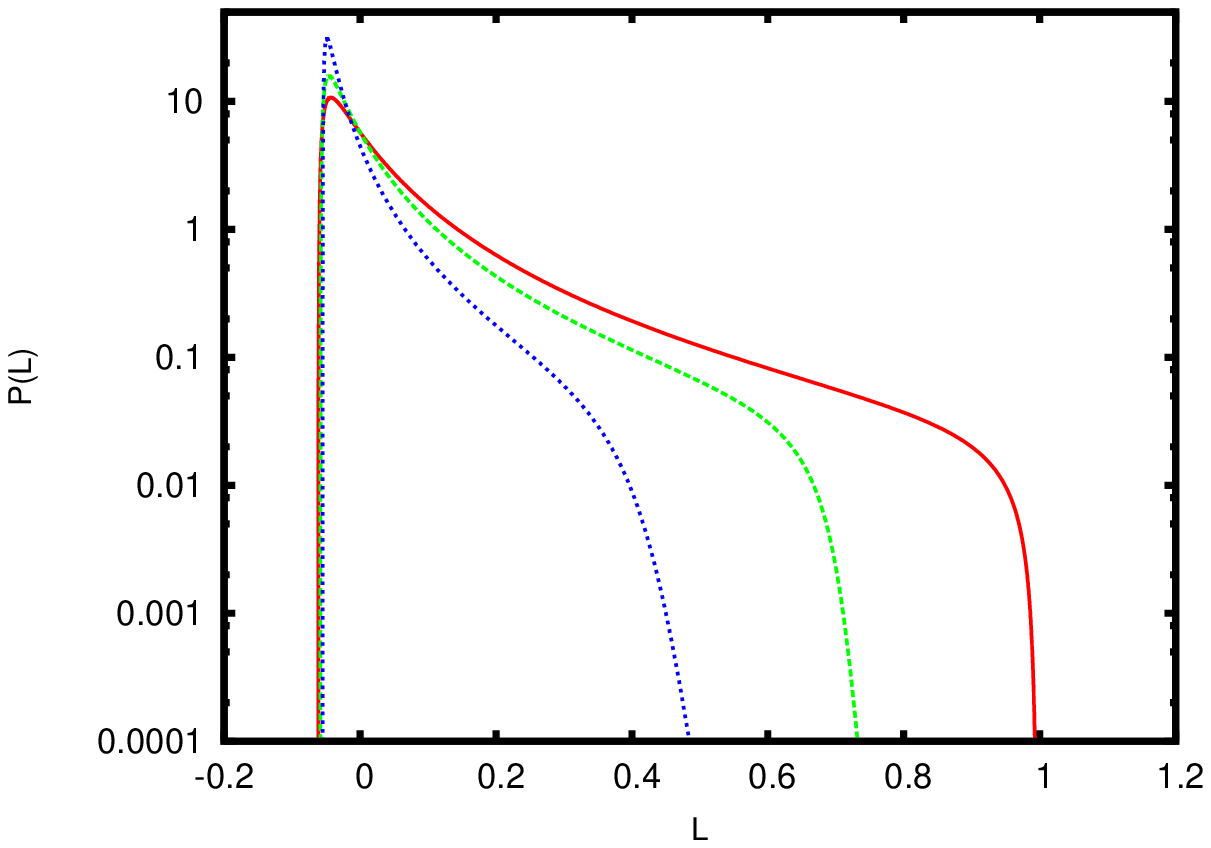,width=8.0cm}}
\put(90,0){\epsfig{file=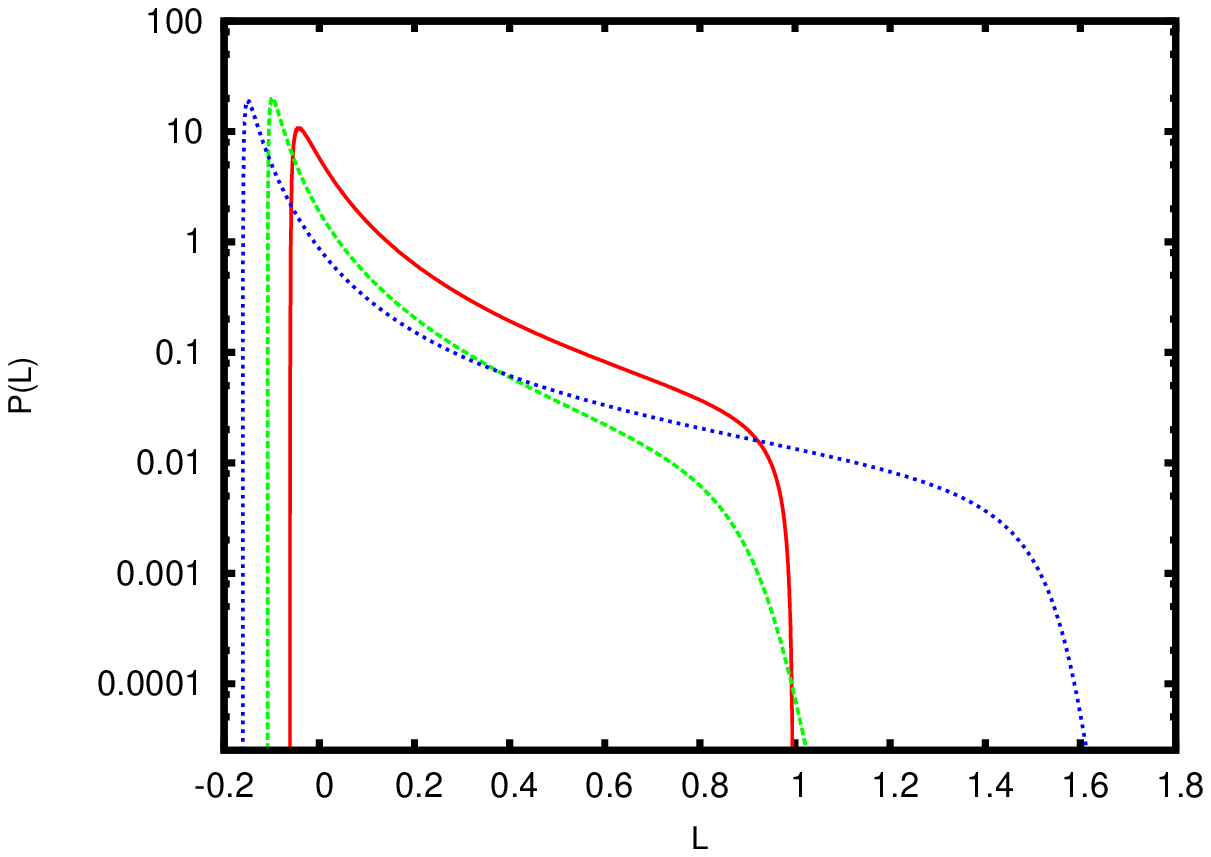,width=8.0cm}}
\end{picture}
\caption{Left: Loss distributions for the baseline-scenario ($\bm S_0$, red full line), compared to scenarios where either one third ($\bm S_5$, green long-dashed) or two thirds ($\bm S_6$, blue short-dashed) of the average exposure to firms are hedged by CDS, but now with protection provided by the {\em insurance sector\/}, which reduces the tail risk of large losses, but does not completely eliminate it. Right: Here the baseline-scenario ($\bm S_0$, red full line) is compared with situations where banks have completely hedged their exposures through CDS  and used this to either double ($\bm S_7$, green long-dashed) or triple ($\bm S_8$, blue short-dashed) the size of their loan books.}
\label{fig:Res2}
\end{figure}

In the first panel of Fig. \ref{fig:Res2}, we compare the loss distribution for the baseline-scenario with a situation where either one third or two thirds of the baseline-exposure to firms are hedged by CDS, but now with protection provided by the {\em insurance sector\/}, thereby supposedly deflecting {\em all\/} potential losses from the hedged component 
into the insurance sector. The idea of CDS being that they would provide protection against losses, one might thus naively expect that hedging a fraction $f_h$ of original exposures with insurers would lead to a saturation of maximal losses  {\em inside the banking sector\/} at the residual fraction $1-f_h$ of exposures, which remains unhedged. The figure clearly shows that this naive expectation is incorrect: for both, $f_h=1/3$, and $f_h=2/3$, we find that the loss distribution gives non-negligible weight to losses {\em in excess\/} of the naively expected saturation level, which would be at a fraction $1-f_h$ of the baseline exposures. This is due to the fact that under very adverse economic conditions the providers of protection may eventually also collapse, rendering loans effectively unhedged.

The second panel compares the baseline scenario with situations where banks use {\em complete hedging\/} via CDS to double or triple the size of their loan books, hedging half the original exposure within the banking sector, and the remainder with insurers. The figure clearly reveals an incentive for such a strategy, namely typical earnings that are increased in proportion to the loan-book expansion. However, once the expansion becomes significant, there is a noticeable risk of incurring significantly higher losses within the banking sector as well, despite the fact that the majority of the exposures is hedged with insurers. Any losses observed in excess of half the baseline exposure would in these scenarios naively be expected to be absorbed by the insurers. The figure once more clearly shows that they are not. Observed excess losses do in fact increase with the expansion of loan-books, as this leads insurers being exposed to larger risk, and thereby to larger default rates in the insurance sector, effectively removing protection and reflecting a sizeable fraction of losses back into the banking sector. While deflecting losses into the insurance sector appears to `effectively' reduce the size of the maximum loss, it does in principle not do this completely. It just {\em considerably\/} reduces the probability of incurring losses close to maximal. In other words, the tail risk of extreme losses cannot be eliminated completely.

Had the corresponding expansion of loan books happened with protection through CDS entirely {\em within the banking sector}, this would --- according to results discussed above --- have exposed the banking sector as a whole to the full risk, with maximum losses at twice or three times the original exposure level, just as in the corresponding unhedged case, as we have seen that hedging through CDS is a zero-sum game and does not affect the loss distribution at system level. 
A loss-distribution for the case of loan-books expanded  by a factor of two is actually displayed in the left panel of Fig. \ref{fig:Res1ab}.

\begin{figure}[ht!] 
\setlength{\unitlength}{1mm}
\begin{picture}(165,60)(0,0)
\put(5,0){\epsfig{file=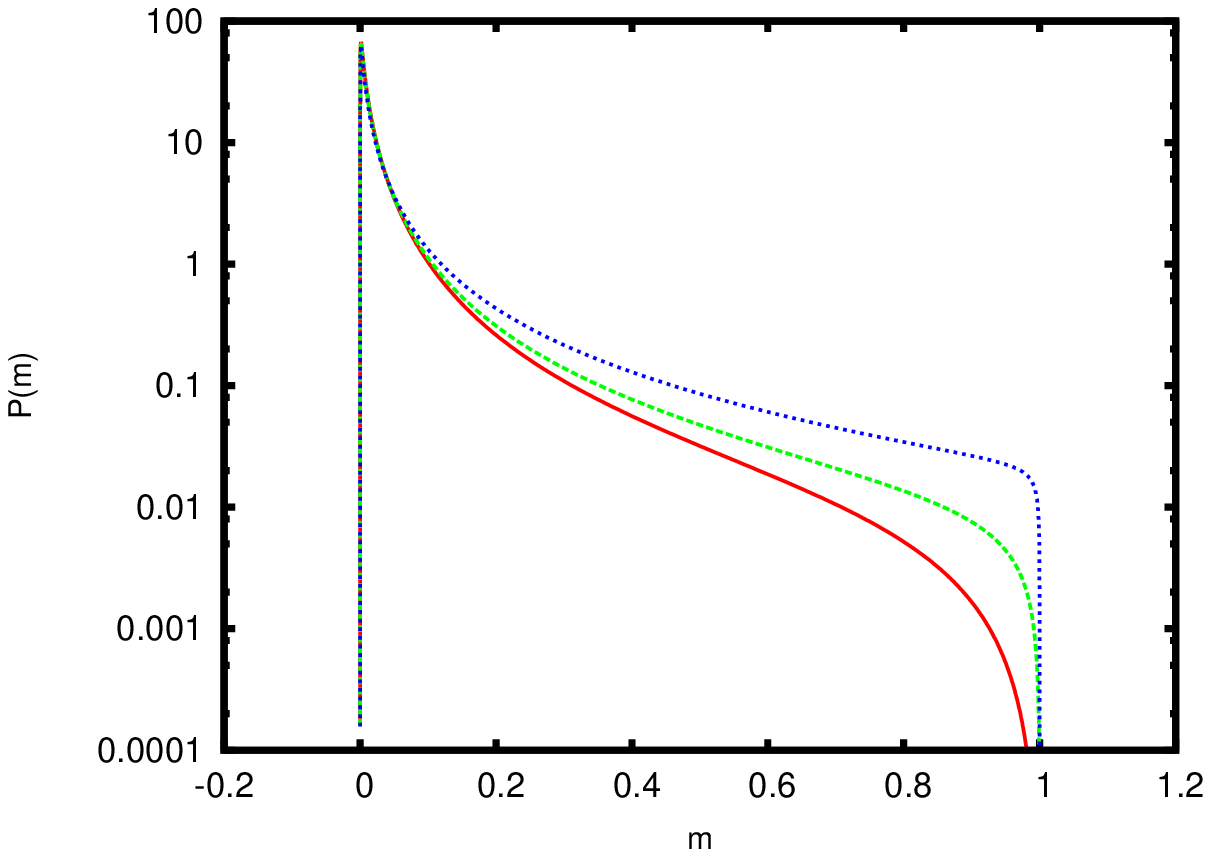,width=8.0cm}}
\put(45.0,28.5){\epsfig{file=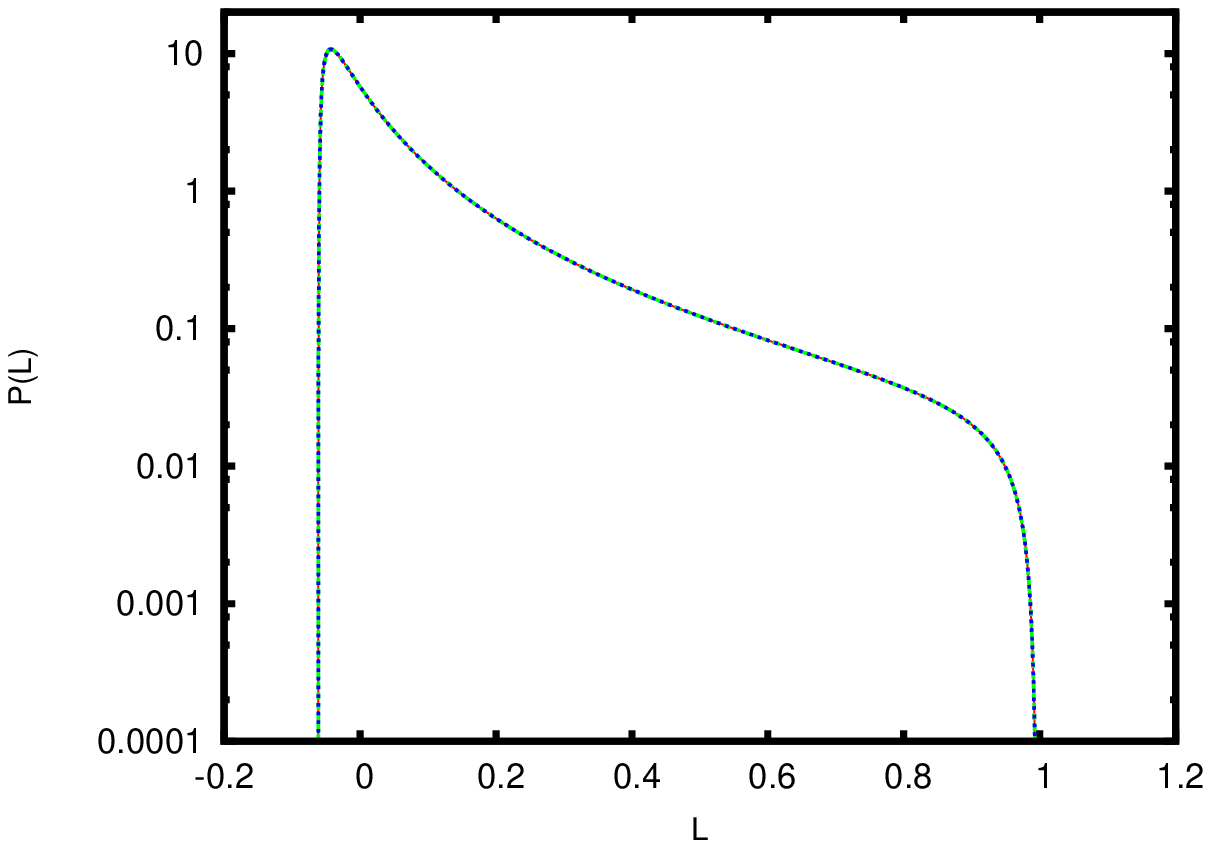,width=3.5cm}}
\put(90,0){\epsfig{file=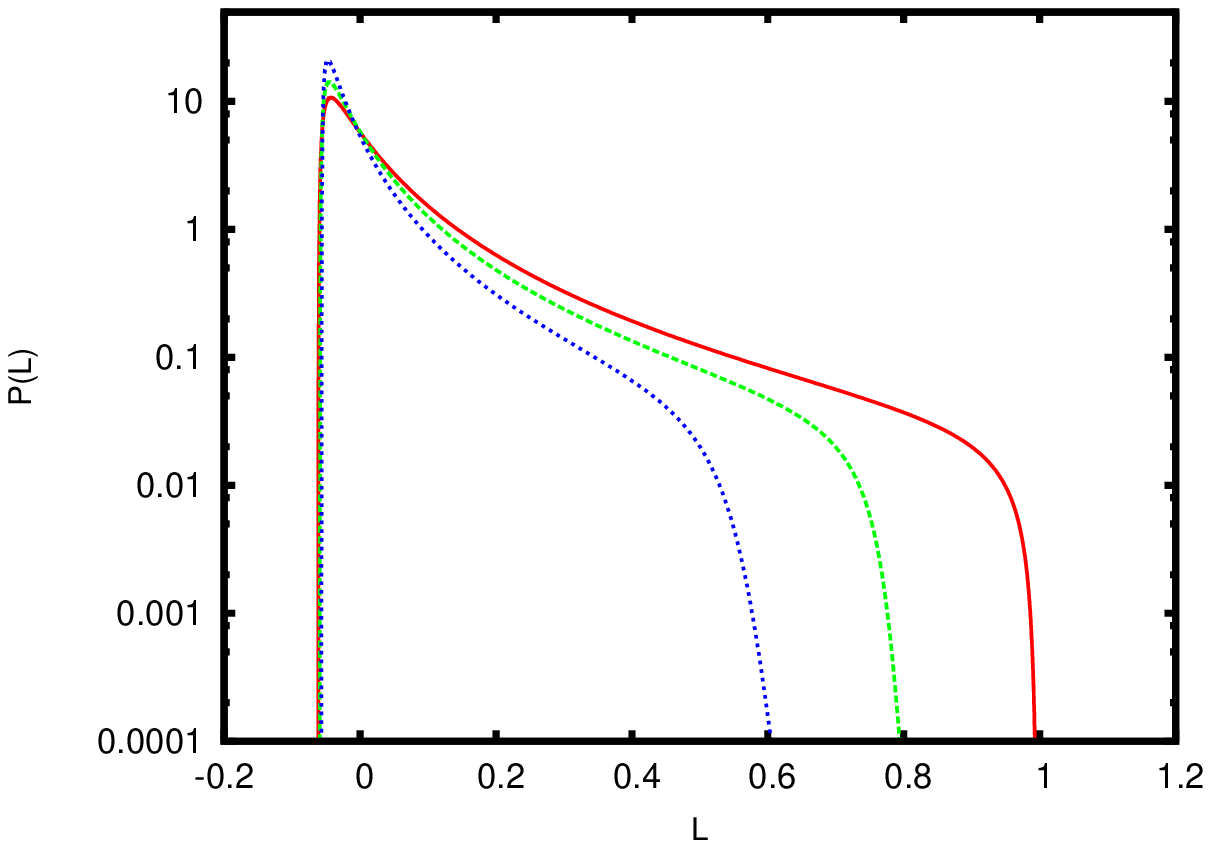,width=8.0cm}}
\end{picture}
\caption{Left: Distribution of the fraction of defaulted banks in the baseline scenario ($\bm S_0$, red full line), compared to situations where speculative CDS of a volume matching that of the base-line exposure  ($\bm S_9$, green long-dashed) or twice that of the base-line exposure ($\bm S_{10}$, blue short-dashed) have been taken out {\em  within the banking sector}. The corresponding loss curves are shown in the inset, and they lie exactly on top of each other, as losses due to speculative CDS, too,  are modelled as a zero-sum game. Right: Loss distributions for the baseline-scenario ($\bm S_0$, red full line), compared to situations with additional speculative CDS taken out in equal measure with banks and insurers with $\bm K^B_{BF}=\bm K^I_{BF}=(0.25,0.125)$ ($\bm S_{11}$, green long-dashed), or  $\bm K^B_{BF}=\bm K^I_{BF}=(0.5,0.25)$ ($\bm S_{12}$, blue short-dashed). }
\label{fig:Res3}
\end{figure}

Let us finally turn to {\em speculative\/} CDS, which nowadays account for approximately two thirds of the total CDS market. It follows from Eqs (\ref{e:Lsb}) and (\ref{e:Lss}) that speculative CDS contracts, too, amount to zero-sum games. Thus, any speculative CDS taken out {\em inside\/} the banking sector only reshuffle losses within the sector, without altering the distribution of losses per node. This is illustrated in the inset of the left panel of Fig. \ref{fig:Res3}. Yet, although speculative CDS amount to zero-sum games and thus do not alter the distribution of losses per node in the sector, there is a marked effect on the distribution of the fraction of defaulted banks, which develops significantly fatter tails at high default fractions once exposures taken on via speculative CDS become sizeable, as shown in the main left panel of Fig. \ref{fig:Res3}. As was the case with CDS used for hedging, this result clearly exhibits the contagious and destabilizing effect of speculative CDS on a systemic level, for which the additional contagion channels created by these contracts must be held responsible.

By contrast, to the extent that speculative CDS are taken out with insurers as protection sellers, these create income for the banking sector, in particular in times of economic stress, which then suppresses the loss distribution at large losses. This is clearly borne out by the results shown in the right panel of Fig. \ref{fig:Res3}. However, this income is provided by inducing equally large losses within the insurance sector which is chosen to take the counter-position, and concentrating on losses within the banking sector therefore paints a picture too rosy for the financial sector as a whole.

\begin{figure}[ht] 
\setlength{\unitlength}{1mm}
\begin{picture}(165,60)(0,0)
\put(5,0){\epsfig{file=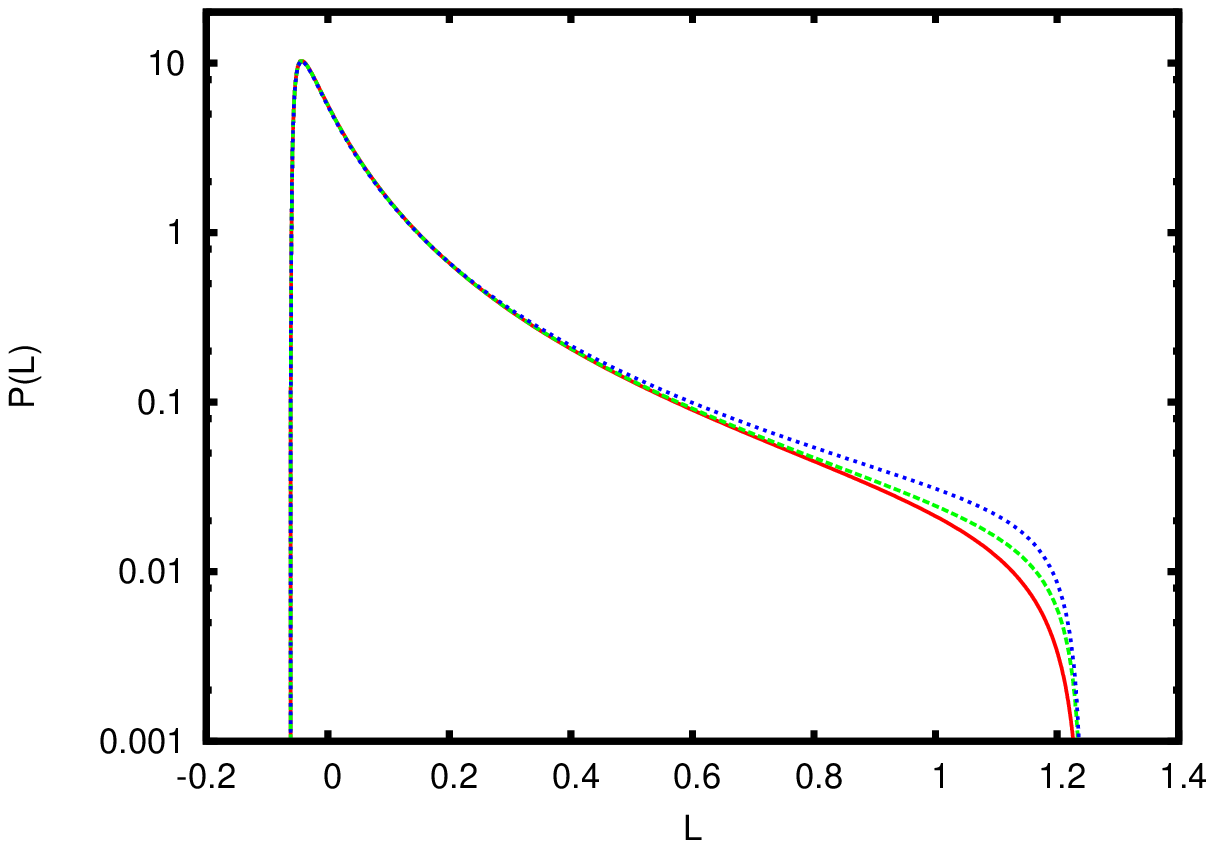,width=8.0cm}}
\put(90,0){\epsfig{file=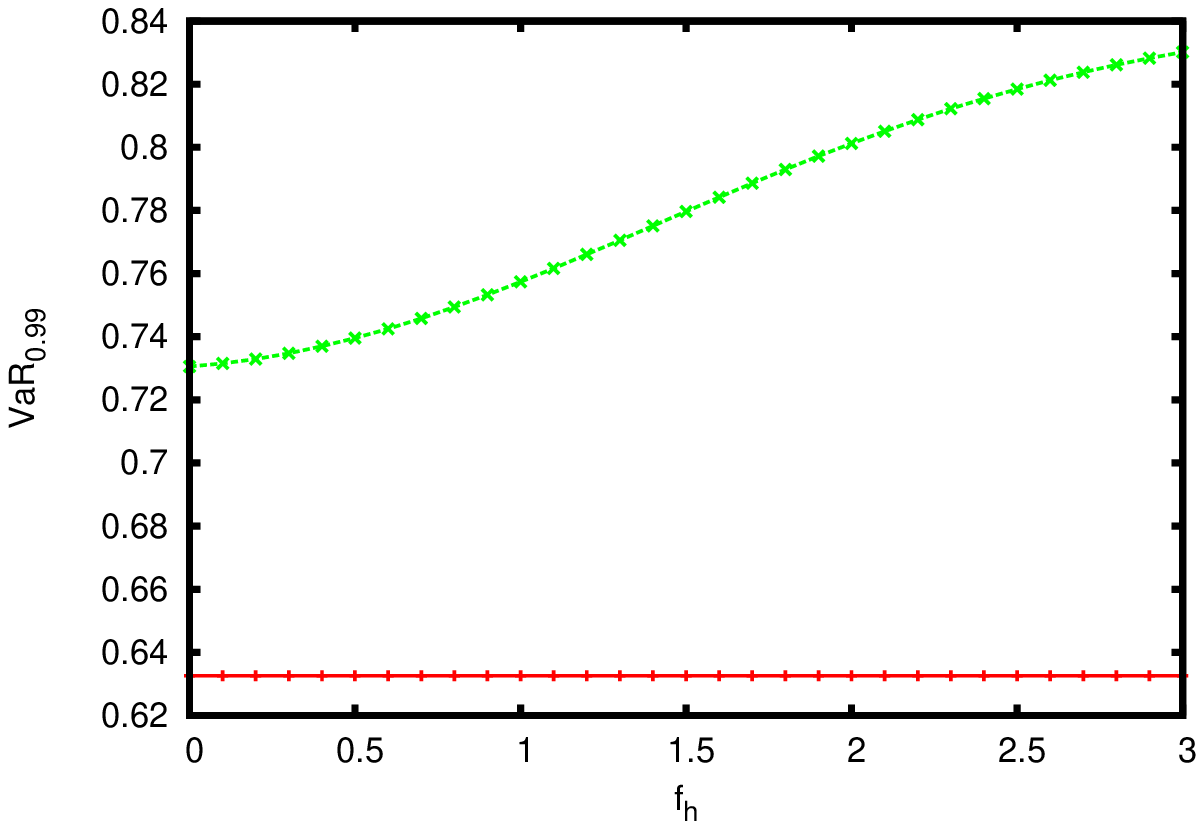,width=8.0cm}}
\end{picture}
\caption{Left: Loss distributions for the modified baseline-scenario ($\bm \tilde S_0$, red full line), compared to similarly modified situations where speculative CDS of a volume matching that of the base-line exposure  ($\bm \tilde S_9$, green long-dashed) or twice that of the base-line exposure ($\bm \tilde S_{10}$, blue short-dashed) have been taken out {\em  within the banking sector}. Despite the fact that (speculative) CDS amount to a zero-sum game, there is now an effect on loss distributions due to direct economic interactions within the banking sector, which are on average cooperative.
Right: Value at risk at the 99\% confidence level as a function of the volume $f_h$ of additional speculative CDS (in 
units of the unhedged baseline-exposure) taken out {\em within the banking sector}. Lower curve: results for $\bm J^{(d)}_{BB} = (0,0.5)$ as in $\bm S_0$. Upper curve: results using the results for $\bm J^{(d)}_{BB} = (0.25,0.5)$ as in $\bm \tilde S_0$.}
\label{fig:Res4}
\end{figure}

It would be fair to argue that the results just shown for the effect of speculative CDS taken out within the banking sector are to some extent counter-intuitive --- to find the probability of large default rates in the banking sector growing with the volume of speculative CDS, without seeing a corresponding effect in loss distributions. In the way these scenarios are set up, this is a consequence of the fact that CDS contracts amount to a zero-sum game, {\em in conjunction\/} with assuming that there will {\em on average\/} not be a negative feedback of defaults within the banking sector by choosing $\bm J^{(d)}_{BB}=(0,0.5)$. This could, however, be regarded as a too optimistic assumption. Indeed, if one were to assume that wealth positions of banks will be on average negatively affected by other defaults in the banking sector, this picture changes as demonstrated in the left panel of Fig. \ref{fig:Res4}. Despite the fact that the money-streams associated with CDS contracts still amount to a zero-sum game, the additional defaults created by contagion through these speculative CDSs now induce additional losses per node within the banking sector, as a default of a bank is now assumed {\em on average\/} have a detrimental effect on the wealth position of other banks via direct economic interactions, by taking $\bm J^{(d)}_{BB}=(0.25,0.5)$. The right  panel investigates this effect more systematically by looking at how the value at risk (VaR) of the loss distribution for the banking sector evolves with the volume of speculative CDS taken out inside the  sector. Measuring this volume in units of the unhedged baseline-exposure, we put $\bm K^{B}_{BF} =f_h (1,0.5)$. The lower curve takes the baseline-scenario as its starting point, for which there is on average {\em no\/} negative feedback of defaults within the banking sector, i.e. $\bm J^{(d)}_{BB}=(0,0.5)$. By contrast, we take the modified base-line-scenario {\em with\/} on average negative feedback of defaults inside the banking sector, i.e. $\bm J^{(d)}_{BB} = (0.25,0.5)$, as the starting point for the upper-curve. The points at $f_h=1$ and $f_h=2$ on the lower curve would represent scenarios  $\bm S_9$ and $\bm S_{10}$, respectively; on the upper curve they represent the corresponding modified scenarios $\bm \tilde S_9$ and $\bm \tilde S_{10}$. For the lower curve without negative feedback, speculative CDS taken out inside the banking sector do not change the loss-distribution and hence leave the VaR invariant. The upper curve on the other hand shows that, as soon as there is negative feedback of defaults through direct interactions inside the banking sector, the increased probability of large default rates created by speculative CDS exposures does modify the loss distribution for the sector, leading to a substantial rise in the VaR.

Analogous negative feedback effects could be created by assuming that firms will on average be adversely affected by defaults in the banking sector by choosing $\bm J^{(d)}_{FB} =(\bar J^{(d)}_{FB}, J^{(d)}_{FB})$ with $\bar J^{(d)}_{FB}>0$. This would then lead to higher default rates in the firm sector, which in turn would create further losses in the financial sector through additional loans that will have to be written off. In the scenarios looked at within the present study we have so far not included this kind of negative feedback by taking $\bm J^{(d)}_{FB} =(0.0)$, though post-crisis debates revolving around the reluctance of banks to lend to firms could well justify introducing it.

\section{Summary and Discussion}
\label{sec:Sum}

In summary, we have proposed a model which allows us to analyze contagion dynamics and {\em systemic risk\/} in networks of financial dependencies which include exposures created by CDS contracts as additional contagion channels. The model generalizes earlier work \cite{NeuKu04, HaKu06, HaKu09} on the influence of economic interactions on credit risk. The main additional features are {\bf (i)} the presence of three-particle interactions, and {\bf (ii)} the non-Markovian nature of contagion dynamics in systems with CDS contracts.

We have investigated a synthetic version of the problem using a stochastic setting in which we assume weighted Erd\"os-Renyi random graph structures \cite{Bollobas01} describing the interconnected networks of mutual exposures of the economic players, taken to include firms $F$, banks $B$, and insurers $I$, and analogous random graph structures containing hyper-edges linking three network nodes through CDS contracts.

Our main result can be summarized as follows: while CDS can help to reduce losses under normal or favourable conditions -- the terms `normal' and `favourable' being defined here with respect to the distribution of the macro-economic component $\xi_0$ of the noise -- they cannot completely eliminate the tail risk of very large losses, and may in fact amplify contagion and losses in times of stress, in particular if CDS are used to expand loan portfolios under the (wrong) assumption that hedging would essentially offload any additional risks from banks' balance sheets.

Our results complement those of a recent empirical study of Markose et al. \cite{mark+10}, who reconstruct the network of CDS-exposures of 26 major US banks and an additional so-called outside-entity which represents total CDS exposures of these 26 banks to non-US financial institutions; data used are as recorded in the 4th quarter of 2008. Interactions that would link these exposures to the economic fate of reference entities outside the network of these 26(+1) banks are not included in the study. The network of CDS exposures is constructed using market share as a proxy, and is dominated by the market share of three major players. A series of stress tests performed by these authors which involve simulating consequences of a failure of any of the major players within the reconstructed network of CDS exposures shows severe knock-on effects when payouts on CDS referencing these failed entities are taken into account (Experiment 2 in \cite{mark+10}), clearly demonstrating the destabilizing effect of CDS within that specific system. As Markose et al. look at one specific system, the relative importance of various properties which characterize the financial systems macroscopically is difficult to ascertain. In particular it remains open how far conclusions based on general statistical characteristics of large networks (such as a ``small-world" or clustering property, or the May-Wigner stability criterion \cite{May72}) can carry for a network containing just 27 nodes.

Our investigation takes a more generic approach than \cite{mark+10}. In particular we  also model the effect of economic links between the financial sector and the `real economy', and we look at stability and at distributions of losses and default rates of the entire economic system throughout economic cycles. 

In the present study, we have restricted our attention to looking at loss distributions and distributions of default rates in the banking sector. Indeed, we have used the insurance sector in the present study {\em only\/} to act as a buffer which can absorb losses which are then no longer `seen' in the banking sector, though only to the extent that insurers survive such loss-taking. As it stands, we have not yet modelled insurers as directly connected also to the real economy (the firm sector), which would allow to model more credible income mechanisms for them. We would therefore at this stage regard it as premature to attach much significance to details of loss-distributions inside the insurance sector, which is why we have not included them here.

We are under no illusion to believe that the simple probabilistic networks of direct and CDS exposures as set in the present study up would provide a realistic description of economic dependencies. For the sake of analytic tractability our study has so far been restricted to a situation where the average number of direct mutual exposures and --- in the case of financial institutions --- also the average number of CDS contracts they are involved in is {\em large\/}, entailing that relative fluctuations about these average numbers tend to be small. Future work will have to include introducing more realistic network topologies, exhibiting in particular stronger heterogeneity of economic players. Including these will {\em considerably\/} complicate the mathematical analysis, though it appears that the tools developed in \cite{Hatchett+04} could be adapted to treat situations with more realistic levels of heterogeneity. Numerical simulations would naturally provide another route that could be taken as an alternative. Finally, there are other contagion mechanisms which we have so far omitted; they include the role of capital costs and of losses on SPV credit enhancements induced by breakdown of CDS cover \cite{mark+10}, as well as knock-on effects created by liquidity risks \cite{GaiKap10}.

Yet, in spite of the many simplifying assumptions still contained in the present study, we are confident that our main message concerning the contribution of CDS to systemic risk, which we extract from a {\em comparison of scenarios with and without CDS}, holds tight and will survive (re-)introducing further microscopic detail, through calibration or otherwise.

We believe that our model adds a new dimension to research on credit contagion, and that it could eventually feed into a rational underpinning of an improved regulatory framework for credit derivative markets. Such potential could in particular derive from results of the form indicated in Fig \ref{fig:Res1cd}  which demonstrate that there might be levels of hedging with are optimal with respect to a certain set of risk measures, or those in Fig \ref{fig:Res4} which quantify the destabilizing effect of speculative CDS exposures. However, further work is needed to see whether and in which form these features will survive the introduction of more realistic network topologies and exposure statistics.

\bibliography{/home/kuehn/Proj/MyBib}

\begin{thebibliography}{10}

\bibitem{Haldane09}
A.~G. Haldane.
\newblock {Rethinking the Financial Network}.
\newblock www.bankofengland.co.uk/publications/speeches/2009, 2009.

\bibitem{Turner+10}
A.~Turner, A.~Haldane, P.~Woolley, S.~Wadhwani, A.~Smithers, A.~Large, J.~Kay,
  and M.~Wolf.
\newblock {\em {The Future of Finance: The LSE Report}}.
\newblock LSE, 2010.

\bibitem{ECB09}
{ECB}.
\newblock {Credit Default Swaps and Counterparty Risk}.
\newblock www.ecb.int/pub, 2009.

\bibitem{Haldane11}
A.~G. Haldane and R.~M. May.
\newblock {Systemic Risk in Banking Ecosystems}.
\newblock {\em Nature}, 469:351--355, 2011.

\bibitem{Lando98}
D.~Lando.
\newblock {On Cox Processes and Credit Risky Securities}.
\newblock {\em Rev. Deriv. Res.}, 2:99--120, 1998.

\bibitem{Duffie99}
D.~Duffie and K.~Singleton.
\newblock {Modeling Term Structure Models of Defaultable Bonds}.
\newblock {\em Rev. Fin. Stud.}, 12:197--226, 1999.

\bibitem{Davis99}
M.~Davis and V.~Lo.
\newblock {Infectious Defaults}.
\newblock {\em Quant. Fin.}, 1:111--115, 1999.

\bibitem{Jarrow01}
R.~Jarrow and S.~Turnbull.
\newblock {Pricing Derivatives on Financial Securities Subject to Credit Risk}.
\newblock {\em J. Fin.}, 56:1765--1799, 2001.

\bibitem{Li01}
D.~Li.
\newblock {On Default Correlation: a Copula Function Approach}.
\newblock {\em J. Fixed Inc.}, 9:43--54, 2001.

\bibitem{Frey03}
R.~Frey and J.~Backhaus.
\newblock {Interacting Defaults and Counterparty Risk: a Markovian Approach}.
\newblock Working Paper, University of Leipzig, 2003.

\bibitem{Rogge03}
E.~Rogge and P.~J. Sch\"onbucher.
\newblock {Modelling Dynamic Portfolio Credit Risk}.
\newblock Working Paper. Imperial College, London, ABN Amro Bank, London, and
  ETH, Z\"urich, 2003.

\bibitem{Giesecke04}
K.~Giesecke and S.~Weber.
\newblock {Cyclical Correlation, Credit Contagion, and Portfolio Losses}.
\newblock {\em J. Bank. Fin.}, 28:3009--3036, 2004.

\bibitem{NeuKu04}
P.~Neu and R.~K\"uhn.
\newblock {Credit Risk Enhancement in a Network of Interdependent Firms}.
\newblock {\em Physica A}, 342:639--655, 2004.

\bibitem{HaKu06}
J.~P.~L. Hatchett and R.~K\"uhn.
\newblock {Effects of Economic Interactions on Credit Risk}.
\newblock {\em J. Phys. A}, 39:2231--2251, 2006.

\bibitem{Egloff07}
D.~Egloff, M.~Leippold, and P.~Vannini.
\newblock {A Simple Model of Credit Contagion}.
\newblock {\em J. Bank. Fin.}, 31:2475--2492, 2007.

\bibitem{HaKu09}
J.~P.~L. Hatchett and R.~K\"uhn.
\newblock {Credit Contagion and Credit Risk}.
\newblock {\em Quant. Fin.}, 9:373--382, 2009.

\bibitem{Keenan00}
S.~Keenan.
\newblock {Historical Default Rates of Corporate Bond Issuers, 1920-1999}.
\newblock Moody's Investor Services, 2000.

\bibitem{Das07}
S.~R. Das, D.~Duffie, N.~Kapdia, and L.Saita.
\newblock {Common Failings: How Corporate Defaults Are Correlated}.
\newblock {\em J. Fin.}, 62:93--117, 2007.

\bibitem{Duff+09}
D.~Duffie, A.~Eckner, G.Horel, and L.~Saita.
\newblock {Frailty Correlated Default}.
\newblock {\em J. Fin.}, 64:2089--2123, 2009.

\bibitem{Az+11}
S.~Azizpour, K.~Giesecke, and G.~Schwenkler.
\newblock {Exploring the Sources of Default Clustering}, 2011.

\bibitem{Hull00}
J.~Hull and A.~White.
\newblock {Valuing Credit Default Swaps I: No Counterparty Default Risk}.
\newblock {\em J. Deriv}, 8(1):29--40, 2000.

\bibitem{Hull01}
J.~Hull and A.~White.
\newblock {Valuing Credit Default Swaps II: Modeling Default}.
\newblock {\em J. Deriv}, 8(3):12--22, 2001.

\bibitem{Haworth07}
H.~Haworth and C.~Reisinger.
\newblock {Modeling Basket Credit Defaults Swaps with Default Contagion}.
\newblock {\em J. Cred. Risk}, 3:31--67, 2007.

\bibitem{Haworth08}
H.~Haworth, C.~Reisinger, and W.~Shaw.
\newblock {Modelling Bonds and Credit Default Swaps Using a Structural Model
  with Contagion}.
\newblock {\em Quant. Fin.}, 8:669--680, 2008.

\bibitem{Errais10}
E.~Errais, K.~Giesecke, and L.~Goldberg.
\newblock {Affine Point Processes and Portfolio Credit Risk}.
\newblock {\em Siam J. Financial Math.}, 1:642--665, 2010.

\bibitem{Frey08}
R.~Frey and J.~Backhaus.
\newblock {Pricing and Hedging of Portfolio Credit Derivatives with Interacting
  Default Intensities}.
\newblock {\em Int. J. of Theor. Appl.Fin.}, 11:611--634, 2008.

\bibitem{Brigo09}
D.~Brigo and K.~Chourdakis.
\newblock {Counterparty Risk for Credit Default Swaps: Impact of Spread
  Volatility and Default Correlation}.
\newblock {\em Int. J. Theor. Appl. Fin.}, 12:1007--1026, 2009.

\bibitem{Frey10}
R.~Frey and J.~Backhaus.
\newblock {Dynamic Hedging of Synthetic CDO-Tranches with Spread- and Contagion
  Risk}.
\newblock {\em J. Econ. Dyn. Contr.}, 34:710--724, 2010.

\bibitem{Cous+11}
A.~Cousin, M.~Jeanblanc, and J.-P. Laurent.
\newblock {Hedging CDO Tranches in a Markovian Environment}.
\newblock In R.~A. Carmona, E.~Cinlar, I.~Ekeland, and E.~Jouini, editors, {\em
  {Paris-Princeton Lectures on Mathematical Finance 2010}}, pages 1--62.
  Springer, Berlin, Heidelberg, 2011.

\bibitem{Jorion07}
P.~Jorion and G.~Zhang.
\newblock {Good and Bad Credit Contagion: Evidence From Credit Default Swaps}.
\newblock {\em J. Fin. Econ.}, 84:860--883, 2007.

\bibitem{mark+10}
S.~Markose, S.~Giansante, M.~Gatkowski, and A.~R. Shaghaghi.
\newblock {Too Interconnected To Fail: Financial Contagion and Systemic Risk In
  Network Model of CDS and Other Credit Enhancement Obligations of US Banks}.
\newblock {Presentation given at the ECB Workshop on {\em Recent Advances in
  Modelling Systemic Risk Using network}, October 2009,COMISEF Working Papers
  Series WPS-033 21/04/2010}, 2010.

\bibitem{Elsinger+06}
H.~Elsinger, A.~Lehar, and M.~Summer.
\newblock {Risk Assessment for Banking Systems}.
\newblock {\em Management Science}, 52:1301--1314, 2006.

\bibitem{Schwei+09}
F.~Schweitzer, G.~Fagiolo, D.~Sornette, F.~Vega-Redondo, A.~Vespignani, and
  D.~R. White.
\newblock {Economic Networks: The New Challenges}.
\newblock {\em Science}, 325:422--425, 2009.

\bibitem{Cont+09}
R.~Cont, E.~Bastos~e Santos, and A.~Moussa.
\newblock {Network Structure and Systemic Risk in Banking Systems}.
\newblock working paper, SSRN-1733528, 2009.

\bibitem{GaiKap10}
P.~Gai an~S.~Kapadia.
\newblock {Contagion in financial Networks}.
\newblock {\em Proc. R. Soc. A}, 466:2401--2423, 2010.

\bibitem{Amini+10}
H.~Amini amd R.~Cont and A.~Minca.
\newblock {Stress Testing the Resilience of Financial Networks}.
\newblock {\em preprint, SSRN-1731292, to appear in Int. J. Theor. Appl. Fin},
  15, 2010.

\bibitem{Caccioli+09}
F.~Caccioli, P.~Vivo, and M.~Marsili.
\newblock {Eroding Market Stability by Proliferation of Financial Instruments}.
\newblock {\em Eur. Phys. J. B}, 71:467--479, 2009.

\bibitem{ISDA10}
ISDA.
\newblock {ISDA Market Survey Annual Data}.
\newblock www.isda.org, 2010.

\bibitem{Derrida87}
B.~Derrida, E.~Gardner, and A.~Zippelius.
\newblock An exactly soluble asymmetric neural network model.
\newblock {\em Europhys. Lett.}, 4:167--173, 1987.

\bibitem{Dominicis78}
C.~De Dominics.
\newblock {Dynamics as Substitute for Replicas in Systems with Quenched Random
  Impurities}.
\newblock {\em Phys. Rev. B}, 18:4913--4919, 1978.

\bibitem{Bollobas01}
B.~Bollob\`as.
\newblock {\em {Random Graphs}}.
\newblock Cambridge Univ. Press, Cambridge, 2001.

\bibitem{Basel05}
Basel~Committee on~Banking~Supervision.
\newblock {International Convergence of Capital Measurement and Capital
  Standards: A Revised Framework. Part 2, Section III: Credit Risk - The
  Internal Ratings Based Approach}.
\newblock www.bis.org, 2005.

\bibitem{May72}
R.~M. May.
\newblock {Will a Large Complex System be Stable?}
\newblock {\em Nature}, 238:413--414, 1972.

\bibitem{Hatchett+04}
J.~P.~L. Hatchett, B.~Wemmenhove, I.~P\'{e}rez~Castillo, T.~Nikoletopoulos,
  N.~S. Skantzos, and A.~C.~C. Coolen.
\newblock {Parallel Dynamics of Disordered Ising Spin Systems on Finitely
  Connected Random Graphs}.
\newblock {\em J. Phys. A}, 37:6201--6220, 2004.

\end{thebibliography}

\appendix
\section{Correlations Needed for Loss Variances}
\label{sec:AppCorr}
In this appendix we collect the calculation of two-time correlations that are needed for the calculations of variances for the various loss types, as we have encountered them in the calculation of the interest-interest contribution (\ref{e:eps-eps}) and the contagion interest contribution (\ref{e:c-eps}) to the loss variance for unhedged loans. We will derive only contributions of individual counter-parties, or pairs of counter-parties, where necessary for losses that involve CDS-contracts.

Key identities that will be used repeatedly are
\be
n_{j,\tau}n_{j,\tau'}=n_{j,\tau^<}~~,~~~~ (1-n_{j,\tau})(1-n_{j,\tau'}) =
(1-n_{j,\tau^>})
\label{e:corr}
\ee
with $\tau^<={\rm min}\{\tau,\tau'\}$, and $\tau^>={\rm max}\{\tau,\tau'\}$

We begin with the correlations required for the interest-interest, fee-fee, and interest-fee contributions as they are independent of loss-type in which they occur

The interest-interest contribution to variances requires the evaluation of
\bea
V_{\epsilon-\epsilon}=\sum_{\tau=1}^t \sum_{\tau'=1}^t \overline{\langle
\epsilon_{ij\tau}\epsilon_{ij\tau'}\rangle}
&=&\epsilon_{ss'}^2 \sum_{\tau,\tau'=1}^t (1+\epsilon_{ss'})^{\tau-1}
(1+\epsilon_{ss'})^{\tau'-1}\overline{\langle(1-n_{j,\tau})(1-n_{j,\tau'})
\rangle}\nn
\eea
Using (\ref{e:corr}) we can rewrite this as
\bea
V_{\epsilon-\epsilon} &=& \epsilon_{ss'}^2 \sum_{\tau=1}^t (1+\epsilon_{ss'})^{\tau-1}
\Bigg[\sum_{\tau'=1}^{\tau-1} (1+\epsilon_{ss'})^{\tau'-1}\overline{\langle(1-n_{j,\tau})\rangle}
+\sum_{\tau'=\tau}^t(1+\epsilon_{ss'})^{\tau'-1}\overline{\langle(1-n_{j,\tau'})\rangle}\Bigg]
\nn\\
&=& \epsilon_{ss'} \sum_{\tau=1}^t(1+\epsilon_{ss'})^{\tau-1} \Big[(1+\epsilon_{ss'})^{\tau-1}-1\Big]\overline{\langle(1-n_{j,\tau})\rangle}\nn\\
& & + \epsilon_{ss'} \sum_{\tau'=1}^t  (1+\epsilon_{ss'})^{\tau'-1} \,\Big[(1+\epsilon_{ss'})^{\tau'}-1\Big] \overline{\langle(1-n_{j,\tau'})\rangle}\nn
\eea
Exchanging $\tau\leftrightarrow \tau'$ in the last line allows to combine the result into
\be
V_{\epsilon-\epsilon} = \epsilon_{ss'}\sum_{\tau =1}^t (1+\epsilon_{ss'})^{\tau-1}
\Big[(1+\epsilon_{ss'})^{\tau-1}+ (1+\epsilon_{ss'})^{\tau}-2\Big]\,
\overline{\langle (1-n_{j,\tau})\rangle}
\ee

Next, consider the fee-fee contribution
\bea
V_{f-f} = \sum_{\tau=1}^t \sum_{\tau'=1}^t \overline{\langle f_{ij,\tau}^k 
f_{ij,\tau'}^k \rangle}
&=& f_0^2 \sum_{\tau=1}^t \sum_{\tau'=1}^t \overline{\langle (1-n_{i,\tau})
(1-n_{j,\tau})(1-n_{k,\tau}) (1-n_{i,\tau'})(1-n_{j,\tau'}) (1-n_{k,\tau'})
\rangle}\nn\\
& & + f^2 \sum_{\tau=1}^t \overline{\langle (1-n_{i,\tau})(1-n_{j,\tau})
(1-n_{k,\tau})\rangle}\nn
\eea
For the dynamics we need it only conditioned on $n_{i,t}=0$, hence $n_{i,\tau}=0$, $t\le \tau$. Using (\ref{e:corr}) we get
\be
V_{f-f}\Big|_{n_{i,t}=0}= \sum_{\tau=1}^t [f_0^2 (2\tau-1) + f^2] \,
\overline{\langle(1-n_{j,\tau})(1-n_{k,\tau})\rangle}
\ee

The fee-interest contribution finally (for the dynamics needed only conditioned on $n_{i,t}=0$, i.e. the protection buyer in (hb)/(sb) positions or the protection  seller in (hs)/(ss) positions being alive) is evaluated along similar lines, giving
\be
V_{f-\epsilon} = f_0 \sum_{\tau=1}^t\Bigg[\big[(1+\epsilon_{ss'})^{\tau-1}-1 \big]
\overline{\langle (1-n_{j,\tau}) (1-n_{k,\tau})\rangle} + \sum_{\tau'=1}^\tau
(1+\epsilon_{ss'})^{\tau-1}\overline{\langle (1-n_{j,\tau}) (1-n_{k,\tau'})\rangle}\Bigg]
\ee
\noindent
The remainder is specific to individual loss positions.

\noindent
For {\bf unhedged loans (u)}\\
--- the contagion-interest contribution to the variance requires
\be
V^{(u)}_{c-\epsilon}= \epsilon_{ss'} \sum_{\tau=1}^t (1+\epsilon_{ss'})^{\tau-1} 
\overline{\langle n_{j,t}(1-n_{j,\tau})\rangle}= 
\epsilon_{ss'} \sum_{\tau=1}^t (1+\epsilon_{ss'})^{\tau-1} 
\overline{\langle (n_{j,t}-n_{j,\tau})\rangle}
\ee
For the {\bf protection buyer position (hb)} in {\bf hedged loans }\\
--- the contagion-contagion contribution to the variance requires
\bea
V^{(hb)}_{c-c}&=&\sum_{\tau,\tau'=1}^t \overline{\langle (n_{j,\tau}-n_{j,\tau-1})
(n_{j,\tau'}-n_{j,\tau'-1}) n_{k,\tau}n_{k,\tau'}\rangle}\nn\\
&=&\sum_{\tau=1}^t\sum_{\tau'=1}^\tau \overline{\langle n_{k,\tau'}
\Big[n_{j,\tau} n_{j,\tau'}+ n_{j,\tau-1}n_{j,\tau'-1}-n_{j,\tau}n_{j,\tau'-1}
-n_{j,\tau'}n_{j,\tau-1}\Big]
\rangle}\nn\\
& & +\sum_{\tau=1}^{t}\sum_{\tau'=\tau+1}^{t}\overline{\langle n_{k,\tau}
\Big[n_{j,\tau} n_{j,\tau'}+ n_{j,\tau-1}n_{j,\tau'-1}-n_{j,\tau}n_{j,\tau'-1}
-n_{j,\tau'}n_{j,\tau-1}\Big]\rangle}\nn
\eea
Using (\ref{e:corr}) we find that all terms but the $\tau'=\tau$ contribution in the square bracket of the first sum cancel, and similarly all terms in the square brackets of the second sum, giving
\be
V^{(hb)}_{c-c}= \sum_{\tau=1}^t \overline{\langle (n_{j,\tau}-n_{j,\tau-1}) n_{k,\tau}\rangle}
\ee
--- the contagion-fee contribution is of the form
\bea
V^{(hb)}_{c-f}\Big|_{n_{i,t}=0}&=& \sum_{\tau,\tau'=1}^t
 \overline{\langle (n_{j,\tau}-n_{j,\tau-1}) n_{k,\tau} 
f_{ij,\tau'}^k\rangle}
= f_0 \sum_{\tau,\tau'=1}^t \overline{\langle (n_{j,\tau}-n_{j,\tau-1})n_{k,\tau}
(1-n_{j,\tau'}) (1-n_{k,\tau'})\rangle}\nn\\
&=& f_0\sum_{\tau=1}^t \sum_{\tau'=1}^t \Big[\overline{\langle(n_{j,\tau}-n_{j,\tau}n_{j,\tau'})
(n_{k,\tau}-n_{k,\tau}n_{k,\tau'})\rangle}- \overline{\langle(n_{j,\tau-1}-n_{j,\tau-1}n_{j,\tau'}) 
(n_{k,\tau}-n_{k,\tau}n_{k,\tau'})\rangle}\Big]\nn\\
&=& f_0\sum_{\tau=1}^t\sum_{\tau'=1}^\tau \, \overline{\langle(n_{j,\tau}-n_{j,\tau'})
(n_{k,\tau}-n_{k,\tau'})\rangle} -f_0\sum_{\tau=1}^t\sum_{\tau'=1}^{\tau-1} \,
\overline{\langle(n_{j,\tau-1}-n_{j,\tau'}) (n_{k,\tau}-n_{k,\tau'})\rangle}\ ,\nn
\eea
where we exploit the fact that there are no $\tau'>\tau$ contributions, and further that the $\tau'$ summation in the first sum in the last line also  effectively extends only to $\tau-1$, hence
\be
V^{(hb)}_{c-f}\Big|_{n_{i,t}=0}= f_0 \sum_{\tau=1}^t\sum_{\tau'=1}^{\tau-1} \,
\overline{\langle(n_{j,\tau}-n_{j,\tau-1}) (n_{k,\tau}-n_{k,\tau'})\rangle}
\ee
--- the contagion-interest contribution is
\bea
V^{(hb)}_{c-\epsilon} &=& \sum_{\tau,\tau'=1}^t  \overline{\langle (n_{j,\tau}-n_{j,\tau-1}) 
n_{k,\tau} \epsilon_{ij,\tau'}\rangle} \nn\\
&=& \epsilon_{ss'} \sum_{\tau=1}^t \sum_{\tau'=1}^t 
(1+\epsilon_{ss'})^{\tau'-1} \overline{\langle (n_{j,\tau}-n_{j,\tau-1}) n_{k,\tau}
(1-n_{j,\tau'})\rangle}\nn\\
&=&\epsilon_{ss'} \sum_{\tau=1}^t \sum_{\tau'=1}^{\tau-1} (1+\epsilon_{ss'})^{\tau'-1}
\overline{\langle \Big[n_{j,\tau} -n_{j,\tau-1} -n_{j,\tau}n_{j,\tau'}
+n_{j,\tau-1}n_{j,\tau'}\Big]n_{k,\tau}\rangle} \nn\\
&=& \sum_{\tau=1}^t \overline{\langle (n_{j,\tau} -n_{j,\tau-1})n_{k,\tau}\rangle}
\,[(1+\epsilon_{ss'})^{\tau-1}-1]
\eea
Here we have exploited the fact that the contributions inside the square bracket in the next-to-last line vanish for $\tau' \ge \tau$.

\noindent
For the {\bf protection seller position (hs)} in {\bf hedged loans} we obtain simplified versions of those for the {\bf (hb)} position. Results are needed conditioned on $n_{i,t}=0$.\\
--- the contagion-contagion contribution is obtained by omitting the $n_{k,\tau}$ factors in the corresponding (hb) formula, giving
\be
V^{(hs)}_{c-c}\Big|_{n_{i,t}=0}=\sum_{\tau=1}^{t}  \overline{\langle(n_{j,\tau} - 
n_{j,\tau-1})\rangle} = \overline{\langle n_{j,t}\rangle}
\ee
--- the contagion-fee contribution gives
\be
V^{(hs)}_{c-f}\Big|_{n_{i,t}=0}= f_0 \sum_{\tau=1}^t\sum_{\tau'=1}^{\tau-1} \,
\overline{\langle(n_{j,\tau}-n_{j,\tau-1}) (1-n_{k,\tau'})\rangle}
\ee

\noindent
For the {\bf protection buyer position (sb)} in {\bf speculative CDS contracts}\\
--- the contagion-contagion contribution to the variance requires
\bea
V^{(sb)}_{c-c}&=& \sum_{\tau,\tau'=1}^{t}\overline{\langle (n_{j,\tau}-n_{j,\tau-1})
(n_{j,\tau'}-n_{j,\tau'-1})(1-n_{k,\tau})(1-n_{k,\tau'})\rangle}\nn\\
&=&\sum_{\tau=1}^{t}\sum_{\tau'=1}^{\tau}\overline{\langle (1-n_{k,\tau})
\Big[n_{j,\tau} n_{j,\tau'}+ n_{j,\tau-1}n_{j,\tau'-1}-n_{j,\tau}n_{j,\tau'-1}
-n_{j,\tau'}n_{j,\tau-1}\Big]\rangle}\nn\\
& & +\sum_{\tau=1}^{t}\sum_{\tau'=\tau+1}^{t}\overline{\langle (1-n_{k,\tau'})
\Big[n_{j,\tau} n_{j,\tau'}+ n_{j,\tau-1}n_{j,\tau'-1}-n_{j,\tau}n_{j,\tau'-1}
-n_{j,\tau'}n_{j,\tau-1}\Big]\rangle}\nn
\eea
Using (\ref{e:corr}) we find that all terms but the $\tau'=\tau$ contribution in the square bracket of the first sum cancel, and similarly all terms  in the square brackets of the second sum, giving 
\be
V^{(sb)}_{c-c}= \sum_{\tau=1}^{t}\overline{\langle (n_{j,\tau}-n_{j,\tau-1})(1-n_{k,\tau})
\rangle}
\ee
--- for the contagion-fee contribution we have
\bea
V^{(sb)}_{c-f}\Big|_{n_{i,t}=0}&=& \sum_{\tau,\tau'=1}^{t}\overline{\langle 
(n_{j,\tau}-n_{j,\tau-1})(1-n_{k,\tau}) f_{ij,\tau'}^k\rangle}\nn\\
&=& f_0 \sum_{\tau,\tau'=1}^{t} \overline{\langle (n_{j,\tau}-n_{j,\tau-1})
(1-n_{k,\tau})(1-n_{j,\tau'})((1-n_{k,\tau'})\rangle}\nn\\
&=& f_0\sum_{\tau=1}^{t} \sum_{\tau'=1}^{\tau-1}\overline{\langle (1-n_{k,\tau})\Big[
n_{j,\tau}-n_{j,\tau} n_{j,\tau'}-n_{j,\tau-1}+n_{j,\tau-1}n_{j,\tau'}
\Big]\rangle}\nn\\
& &+ \sum_{\tau=1}^{t} \sum_{\tau'=\tau}^{t}\overline{\langle (1-n_{k,\tau'})\Big[
n_{j,\tau}-n_{j,\tau}n_{j,\tau'}-n_{j,\tau-1}+n_{j,\tau-1}n_{j,\tau'}
\Big]\rangle}\nn\\
&=& f_0\sum_{\tau=1}^{t} \sum_{\tau'=1}^{\tau-1}\overline{\langle (n_{j,\tau}-n_{j,\tau-1})
(1-n_{k,\tau})\rangle} = f_0\sum_{\tau=1}^{t} \overline{\langle (n_{j,\tau}-n_{j,\tau-1})
(1-n_{k,\tau})\rangle}\,(\tau-1)
\eea
where we have exploited the fact that terms in square brackets in the $(\tau'\ge \tau)$-sum cancel.

\noindent
For the {\bf protection seller position (ss)} in {\bf speculative CDS contracts}\\
--- we need a contagion-contagion contribution (for the dynamics only conditioned on the seller being alive); it is a simplified version of the corresponding {\bf(sb)} result obtained by omitting the $(1-n_{k,\tau})$ and $(1-n_{k,\tau'})$ factors.
\bea
V^{(ss)}_{c-c}\Big|_{n_{i,t}=0}&=& \sum_{\tau,\tau'=1}^{t}\overline{\langle 
(n_{j,\tau}-n_{j,\tau-1}) (n_{j,\tau'}-n_{j,\tau'-1})\rangle}\nn\\
&=&\sum_{\tau=1}^{t}\sum_{\tau'=1}^{\tau}\overline{\langle 
\Big[n_{j,\tau} n_{j,\tau'}+ n_{j,\tau-1}n_{j,\tau'-1}-n_{j,\tau}n_{j,\tau'-1}
-n_{j,\tau'}n_{j,\tau-1}\Big]\rangle}\nn\\
& & +\sum_{\tau=1}^{t}\sum_{\tau'=\tau+1}^{t}\overline{\langle 
\Big[n_{j,\tau} n_{j,\tau'}+ n_{j,\tau-1}n_{j,\tau'-1}-n_{j,\tau}n_{j,\tau'-1}
-n_{j,\tau'}n_{j,\tau-1}\Big]\rangle}\nn\\
&=& \sum_{\tau=1}^{t}\overline{\langle (n_{j,\tau}-n_{j,\tau-1}) \rangle}
= \overline{\langle n_{j,t}\rangle}\ ,
\eea
where arguments concerning cancellations in square brackets are as for the corresponding {\bf (sb)} case, and the last step exploits properties of a telescoping sum.\\
--- the contagion-fee contribution to the variance (conditioned on the seller being alive) gives, (exploiting telescoping sums from the start)
\be
V^{(ss)}_{c-f}\Big|_{n_{i,t}=0}= f_0\sum_{\tau=1}^{t} \overline{\langle 
n_{j,t}(1-n_{j,\tau})(1-n_{k,\tau})\rangle}= f_0\sum_{\tau=1}^{t-1} \overline{\langle 
(n_{j,t}-n_{j,\tau})(1-n_{k,\tau})\rangle}
\ee

\end{document}